\documentclass[aip,jcp,reprint,floatfix]{revtex4-2}
\usepackage{amsmath,amssymb}
\usepackage{graphicx}
\usepackage{dcolumn}
\usepackage{bm}
\usepackage{multirow}
\usepackage{hhline}
\usepackage{subfigure}
\usepackage{afterpage}
\usepackage{tabularx}
\usepackage{comment}

\usepackage[utf8]{inputenc}
\usepackage[T1]{fontenc}
\usepackage{mathptmx}

\usepackage{physics}

\usepackage{xcolor}

\usepackage[dvipsnames]{xcolor}

\definecolor{wlipurple}{RGB}{128,0,128}

\graphicspath{{Figures/}}

\begin{document}

\preprint{AIP/123-QED}


\title{Elucidating the Synergetic Interplay between Average Intermolecular Coupling and Coupling Disorder in Short-Time Exciton Transfer}


\author{Siwei Wang}
\affiliation{Department of Chemistry\&Biochemistry, University of Notre Dame, Notre Dame, IN, 46616}

\author{Guangming Liu}
\affiliation{Department of Chemistry\&Biochemistry, University of Notre Dame, Notre Dame, IN, 46616}

\author{Hsing-Ta Chen}
\email{hchen25@nd.edu}
\affiliation{Department of Chemistry\&Biochemistry, University of Notre Dame, Notre Dame, IN, 46616}

\begin{abstract}
Exciton transport in molecular aggregates is a fundamental process governing the performance of organic optoelectronics and light-harvesting systems. While most theoretical studies have emphasized long-time transport behavior, recent advances in ultrafast spectroscopy have brought into focus the short-time regime, in which exciton motion remains ballistic on femtosecond-to-picosecond timescales. In this work, we develop an analytical framework for short-time exciton dynamics in a one-dimensional lattice subject to both on-site energetic (diagonal) disorder and intermolecular coupling (off-diagonal) fluctuations. Utilizing the reciprocal-space analysis, we derive closed-form expressions for the first and second spatial moments considering both localized excitation and moving Gaussian initial conditions. Our analytical and numerical results show that, while the long-time dynamics are influenced by diagonal disorder, the short-time ballistic expansion is governed primarily by off-diagonal disorder. Crucially, we reveal a synergistic interplay between the average intermolecular coupling and the off-diagonal coupling disorder strength, demonstrating that they contribute equivalently to short-time exciton transport. Moreover, we integrate this generic disorder model with a realistic molecular system within the framework of macroscopic quantum electrodynamics, thereby providing a theoretical foundation for characterizing and optimizing ultrafast energy flow of disordered molecular aggregates in complex dielectric media.
\end{abstract}


\maketitle

\section{Introduction}
\label{Sec:Introduction}

Exciton transport in molecular aggregates is a fundamental process governing the efficiency of optoelectronic devices and light-harvesting systems,\cite{brixner2017exciton} and is almost always influenced by various disorder effects.\cite{thomas_disordered_2026,aroeira_coherent_2024} Traditionally, disorder effects in these low-dimensional systems are understood through the lens of their long-time localization behaviors, such as Anderson localization,\cite{anderson1958absence,thouless1970anderson,anderson1980new} where randomness in site energies or coupling strengths acts as a barrier that eventually halts energy propagation\cite{lebedenko2009coherent,haken1973exactly}. However, because excitons possess coherence\cite{spano2010spectral} (often on the scale of femtoseconds to picoseconds\cite{scholes2006excitons}), their short-time dynamics can be sensitive to the initial excitation and follow distinct mechanisms from their long-time localization behavior. Recent studies in low-dimensional lattices have revealed an intriguing phenomenon: while disorder suppresses long-range diffusion, it can accelerate the short-time spread of a wavepacket\cite{cui2023short}. Understanding this transient regime is therefore essential for predicting the exciton diffusion length, which determines whether energy can successfully reach a reaction center or electrode before decay in both natural and artificial light-harvesting systems\cite{scholes2017using,valzelli2024large}.

These dynamics is further complicated by the interplay between different types of disorder. Although on-site energetic disorder (diagonal terms in the excitonic Hamiltonian) is the primary focus in localization studies\cite{valleau2012exciton,zanardini2026disorder}, disorder in exciton coupling (off-diagonal terms in the excitonic Hamiltonian) plays a critical role in molecular aggregates.\cite{theodorou1976extended,soukoulis1981off,soukoulis1982study,troisi_charge_2009,fetherolf_unification_2020,liu_dissecting_2025} This coupling disorder stems from the fact that the exciton transfer integral are highly sensitive to fluctuations in molecular orientation, intermolecular separation, and electromagnetic inhomogeneity\cite{kamalov1996temperature,brixner2017exciton,khazanov_embrace_2023}. Recent advances in momentum-resolved ultrafast spectroscopy have shown that hybrid light–matter excitations can exhibit ballistic motion mediated through their light-like character, yet strongly influenced by their interplay with electron-phonon coupling and other surrounding disorder landscape\cite{balasubrahmaniyam2023enhanced,xu_ultrafast_2023,sandik2025cavity}. These observations highlight the need for a more general description of exciton transport under various disorder effects, specifically one that captures how both energetic and coupling fluctuations modulate the coherent expansion of the exciton wavepacket during its short-time motion.

Numerous analytical studies of exciton wavepacket transfer on quantum lattice models have revealed intriguing phenomena arising from disorder.
Early work by Madhukar and Post provides exact solutions with \emph{dynamic} disorder by employing the reciprocal-space Liouville--von Neumann (LvN) equation \cite{madhukar1977exact} in conjunction with Furutsu–Novikov theorem for stochastic fluctuations.\cite{furutsu1963statistical,novikov1965functionals,athanassoulis2019extensions} In these disordered systems, the mean square displacement of the excitonic wavepacket typically exhibits a crossover from coherent short-time dynamics to incoherent diffusion.
Focusing on the short-time dynamics, it has been shown that disorder can enhance transient diffusivity and ballistic wavepacket expansion.\cite{pereverzev_quantum_2005, tutunnikov_coherent_2023,cui2023short}
Specifically, recent analytical studies focusing on purely diagonal energetic disorder demonstrate that short-time wavepacket growth can become faster on a disordered lattice than on an ordered one.
However, the combined effects of diagonal and off-diagonal \emph{static} disorder are largely unexplored.
Moreover, it remains unclear how these disorder effects manifest in realistic physical systems, where disorder typically stems from molecular orientation, electron-phonon coupling, and electromagnetic inhomogeneity.

In this work, we develop a rigorous analytical framework to investigate short-time exciton dynamics in a one-dimensional lattice model subject to both diagonal and off-diagonal static disorder. Using the reciprocal-space LvN equation, we derive closed-form expressions for the first and second spatial moments, $\langle x(t) \rangle$ and $\langle x^2(t) \rangle$. We further connect this generic disorder model to a realistic physical system by employing the macroscopic quantum electrodynamics (MQED) framework \cite{scheel2008macroscopic,buhmann2013dispersion,wang_robust_2025,hsu_chemistry_2025,liu2026mqed} to treat molecular orientational disorder, thereby grounding our theoretical approach in a tangible physical framework.

The paper is organized as follows. In Section~\ref{Sec:Method}, we focus on a generic disorder model and present the analytical derivation of exciton dynamics in a one-dimensional lattice under the short-time approximation. Specifically, we provide analytical expressions for the first and second spatial moments under two distinct initial conditions: (\textit{i}) localized excitation and (\textit{ii}) a moving Gaussian wavepacket. In Section~\ref{Sec:Discussion}, we numerically validate these analytical results and discuss their dependence on the initial condition. In Section~\ref{Sec:MQED_Example}, we apply the generic disorder model analysis to a realistic exciton transfer system with disordered molecular orientations, where parameters like coupling strengths are determined using MQED. Finally, Section~\ref{Sec:Conclusion} summarizes our findings and outlines directions for future research.


\section{Generic Disorder Model and Theoretical Analysis}
\label{Sec:Method}
\subsection{Hamiltonian and the LvN equation in site representation}
\label{Sec_Hamil_Dyna}

We consider a generic disorder model comprised of two-level emitters, which is described by the system Hamiltonian in the site representation
\begin{align}
    \hat{H} = \frac{1}{2} \sum_{m,n} (\alpha_{mn}+\beta_{mn}) \{ |m\rangle\langle n| + |n\rangle\langle m| \},
\label{Eq:Hamiltonian}
\end{align}
where $|n\rangle$ and $|m\rangle$ represent single-excitation states localized on the $n$-th and $m$-th emitters, respectively. 
Following previous studies\cite{madhukar1977exact,cui2023short}, the parameters $\alpha_{mn}$ and $\beta_{mn}$ correspond to the deterministic and random components of the Hamiltonian matrix elements. In the context of exciton transfer, the on-site transition energy is $E_m = \alpha_{mm} + \beta_{mm}$ and the exciton transfer integral is $V_{mn} = \alpha_{mn} + \beta_{mn}$. 

The random variables $\beta_{mn}$ are assumed to follow a Gaussian distribution with zero mean, $\langle \beta_{mn} \rangle_\mathrm{E} = 0$, where $\langle \cdots \rangle_\mathrm{E}$ denotes the ensemble average. The covariance of the random variables is chosen to be 
\begin{align}
\nonumber
    &\langle \beta_{mn} \beta_{m'n'} \rangle_\mathrm{E} \\
    &= g(m-n)(\delta_{m,m'}\delta_{n,n'} + \delta_{m,n'}\delta_{m',n} - \delta_{m,n}\delta_{m',n'}\delta_{n,n'}),
\label{Eq:Covariance_beta}
\end{align}
where $g(m-n)$ characterizes the strength of the static disorder and depends only on the site separation, reflecting the translational invariance of this linear chain. 
Specifically, we restrict the static disorder to include only diagonal and nearest-neighbor off-diagonal contributions by setting the disorder correlation function as
\begin{align}
\label{Eq:Disorder_Correlation_Simple}
g(m-n)=
\begin{cases}
g_0, & m=n,\\
g_1, & |m-n|=1,\\
0, & \text{otherwise}.
\end{cases}
\end{align}
For simplicity, we neglect long-range deterministic interactions. In addition, the deterministic component of the on-site energy, $\alpha_{nn}=0$, is also set to zero because the Hamiltonian in Eq.~(\ref{Eq:Hamiltonian}) can be transformed into the interaction picture, which removes the uniform on-site energy contribution. Under these assumptions, the deterministic coupling elements are restricted to the nearest neighbors:
\begin{align}
    \alpha_{mn} = \alpha_{nm} = J \delta_{\abs{m-n},1}.
\label{Eq:NN_Coupling_J}
\end{align}

Substituting Eq.~(\ref{Eq:Hamiltonian}) into the Liouville--von Neumann (LvN) equation, $i\hbar \frac{\partial \hat{\rho}(t)}{\partial t} = [\hat{H},\hat{\rho}(t)]$, yields the following equations of motion for the density matrix elements $\rho_{l,r}$ (where $\hat{\rho}(t) = \sum_{l,r} \rho_{l,r}(t) |l\rangle\langle r|$):
\begin{align}
\nonumber
\frac{ \partial \rho_{l,r}(t)}{\partial t}
=&
-\frac{i}{\hbar} J \left[ \rho_{l+1,r}(t) + \rho_{l-1,r}(t) - \rho_{l,r+1}(t) - \rho_{l,r-1}(t)\right] \\
&  - \frac{i}{2\hbar} \sum_{n} \left[ (\beta_{ln} + \beta_{nl})\rho_{n,r}(t) - (\beta_{nr} + \beta_{rn})\rho_{l,n}(t) \right].
\label{Eq:Dynamical_Eq_Site}
\end{align}

\subsection{The LvN equation in reciprocal space}
\label{Sec_Dyna_Reci_Space}

To exploit the translational symmetry of the linear chain, we transform Eq.~(\ref{Eq:Dynamical_Eq_Site}) into the reciprocal space\cite{rips1993stochastic,cui2023short}. We define the discrete Fourier transform in the site basis as $\tilde{f}(k_1,k_2) = \sum_{l,r} e^{-i k_1 l + i k_2 r} f_{l,r}$, where $k_{j} = a \bar{k}_{j}$ denotes the dimensionless momentum, $\bar{k}_j$ is the physical momentum, and $a$ is the lattice constant. The first Brillouin zone is thus defined by $k_{j} \in (-\pi, \pi]$, with the corresponding inverse Fourier transform given by $f_{l,r} = (2\pi)^{-2} \int_{-\pi}^{\pi} \int_{-\pi}^{\pi} dk_1 dk_2 \, e^{ik_1 l - ik_2 r} \tilde{f}(k_1, k_2)$. Applying this transformation to both sides of Eq.~(\ref{Eq:Dynamical_Eq_Site}) yields the equations of motion for the density matrix elements $\tilde{\rho}(k_1,k_2,t)$ in reciprocal space:
\begin{align}
\nonumber
\frac{ \partial \langle \tilde{\rho}(k_1, k_2; t) \rangle_\mathrm{E}}{\partial t}
=&
-\frac{i2J}{\hbar}\big(\cos k_1 - \cos k_2\big)\,
\langle \tilde{\rho}(k_1, k_2; t) \rangle_\mathrm{E}
\\
\nonumber
& -\frac{i}{2\pi\hbar}
\int_{-\pi}^{\pi}\!\!\int_{-\pi}^{\pi} dq\,dq'\;
\Big\langle
\big[
\tilde{\beta}(k_1, q)\delta(q' - k_2) \\
& - \tilde{\beta}(q', k_2)\delta(q - k_1)
\big]\,
\tilde{\rho}(q, q'; t)
\Big\rangle_\mathrm{E},
\label{Eq:Dynamical_Eq_Reci}
\end{align}
where $\delta(q'-k_2)$ is the Dirac delta function, and $\tilde{\beta}(k_1, q)$ denotes the discrete Fourier transform of $\beta_{ln}$. To evaluate the ensemble-averaged terms such as $\langle \tilde{\beta}(k_1,q) \tilde{\rho}(q, q'; t) \rangle_\mathrm{E}$ in Eq.~(\ref{Eq:Dynamical_Eq_Reci}), we utilize the Furutsu-Novikov theorem\cite{furutsu1963statistical,novikov1965functionals,athanassoulis2019extensions}:
\begin{align}
\nonumber
    &\langle\tilde{\beta}(k_1,q) \tilde{\rho}(q,q';t)\rangle_\mathrm{E} \\
    &= \int  \int dx dy \langle \tilde{\beta}(k_1,q)\tilde{\beta}(x,y) \rangle_\mathrm{E} \left\langle \frac{\delta \tilde{\rho}(q,q';t)}{\delta\tilde{\beta}(x,y)} \right\rangle_\mathrm{E}  ,
\label{Eq:Furutsu-Novikov}
\end{align}
to break it down in terms of the covariance of the random variables, $\langle \tilde{\beta}(k_1,q)\tilde{\beta}(x,y) \rangle_\mathrm{E}$, and the functional derivative $\langle \delta \tilde{\rho}(q,q';t) / \delta\tilde{\beta}(x,y) \rangle_\mathrm{E}$.

To evaluate the covariance of the random variables in reciprocal space, we substitute Eqs.~(\ref{Eq:Covariance_beta}) and (\ref{Eq:Disorder_Correlation_Simple}) into the discrete Fourier transform definition:
\begin{align}
\nonumber
    &\langle \tilde{\beta}(k_1,q)\tilde{\beta}(x,y) \rangle_\mathrm{E} = \sum_{l,n} e^{-i k_1 l + i q n} \sum_{m,r} e^{-i x m + i y r }  \langle {\beta}_{ln} {\beta}_{mr} \rangle_\mathrm{E} \\
    & = 2\pi \delta(q+y-k_1-x) \left\{ g_0 + 2g_1 \left[ \cos(k_1+x)+ \cos (k_1-y) \right] \right\}  .
\label{Eq:Covariance_Recip}
\end{align}
To evaluate the ensemble-averaged functional derivative $\langle \delta \tilde{\rho}(q,q';t) / \delta\tilde{\beta}(x,y) \rangle_\mathrm{E}$ in Eq.~(\ref{Eq:Furutsu-Novikov}), we perform a functional differentiation of Eq.~(\ref{Eq:Dynamical_Eq_Reci}) with respect to $\tilde{\beta}(x,y)$. Utilizing the identity $\delta \tilde{\beta}(k_1, q) / \delta \tilde{\beta}(x, y) = \delta(k_1 - x)\delta(q - y)$, we obtain the equation of motion for the response function, $R(k_1,k_2;x,y;t) \equiv \langle \delta \tilde{\rho}(k_1, k_2, t) / \delta \tilde{\beta}(x, y) \rangle_\mathrm{E} $:
\begin{align}
\nonumber
&\frac{\partial R(k_1,k_2;x,y;t)}{\partial t} = -\frac{i2J}{\hbar}\bigl(\cos k_1-\cos k_2\bigr) R(k_1,k_2;x,y;t)
\\
\nonumber
&-\frac{i}{2\pi\hbar}
\Big[ \delta(k_1-x)\tilde{\rho}(y,k_2;t)
- \delta(k_2-y)\tilde{\rho}(k_1,x;t) \Big]
\\
&-\frac{i}{2\pi\hbar}
\int_{-\pi}^{\pi} dz\, \left[ \tilde{\beta}(k_1,z) R(z,k_2;x,y;t)
- \tilde{\beta}(z,k_2) R(k_1,z;x,y;t)
\right].
\label{Eq:response_exact}
\end{align}

In the limit of weak static disorder, the final integral term in Eq.~(\ref{Eq:response_exact}) can be neglected as it represents higher-order corrections in the random variable $\tilde{\beta}$. This approximation effectively closes the hierarchy of equations, enabling a closed-form description of the ensemble-averaged dynamics. Furthermore, assuming the initial state $\tilde{\rho}(k_1,k_2,t=0)$ is prepared independently of the random variables, we can set the initial condition $R(k_1,k_2;x,y;t=0)=0$. Formally integrating the truncated equation of Eq.~(\ref{Eq:response_exact}) then yields the ensemble-averaged functional derivative in Eq.~(\ref{Eq:Furutsu-Novikov}):
\begin{align}
\nonumber
& \left\langle
\frac{\delta \tilde{\rho}(q,q';t)}{\delta \tilde{\beta}(x,y)}
\right\rangle_{\mathrm E}
= -\frac{i}{2\pi\hbar}
\int_0^t d\tau\,
e^{-\frac{i2J}{\hbar}\bigl(\cos q-\cos q'\bigr)(t-\tau) }
\\
& \qquad \quad \times \left[
\delta(q-x)\langle \tilde{\rho}(y,q';\tau)\rangle_{\mathrm E}
- \delta(q'-y)\langle \tilde{\rho}(q,x;\tau)\rangle_{\mathrm E}
\right].
\label{Eq:functional_derivative_final}
\end{align}

By substituting Eqs.~(\ref{Eq:Furutsu-Novikov}), (\ref{Eq:Covariance_Recip}), and (\ref{Eq:functional_derivative_final}) into Eq.~(\ref{Eq:Dynamical_Eq_Reci}), and applying the translation property of integrals for a $2\pi$-periodic function $\tilde{f}(q)$ in reciprocal space ($\int_{-\pi}^{\pi} dq \tilde{f}(q) = \int_{-\pi}^{\pi} dq \tilde{f}(q+k)$), we arrive at the following integro-differential equation:
\begin{align}
\nonumber
&\frac{\partial \langle \tilde{\rho}(k_1,k_2;t) \rangle_\mathrm{E} }{\partial t}
=-\frac{i2J}{\hbar}\big(\cos k_1-\cos k_2\big)\, \langle \tilde{\rho}(k_1,k_2;t) \rangle_\mathrm{E} \\
\nonumber &\quad+\int_0^t d\tau \mathcal{K}_\mathrm{loc}(k_1,k_2;t,\tau)
\langle \tilde{\rho}(k_1,k_2;\tau) \rangle_\mathrm{E} \\
&\quad+\int_0^t d\tau \int_{-\pi}^{\pi} dq \mathcal{K}_\mathrm{scat}(k_1,k_2,q;t,\tau)\langle \tilde{\rho}\big(q+k_1,q+k_2;\tau\big) \rangle_\mathrm{E} .
\label{Eq:Integro-Differ}
\end{align}
Here we define two memory kernels $\mathcal{K}_\mathrm{loc}$ and $\mathcal{K}_\mathrm{scat}$ as induced by the reciprocal-space disorder kernel $\mathcal{G}(x, y, q)$ given by:
\begin{align}
\mathcal{G}(x, y, q) = g_0 + 2g_1 \left[ \cos(q + x + y) + \cos(x - y) \right].
\label{Eq:G_xyq_Definition}
\end{align}
On the one hand, $\mathcal{K}_\mathrm{loc}$ is defined as 
\begin{equation}
\begin{split}
\mathcal{K}_\mathrm{loc}(k_1,k_2;t,\tau)& = - \frac{1}{2\pi\hbar^2}\int_{-\pi}^{\pi} dq \\
&\bigg\{e^{-\frac{i2J}{\hbar}\big(\cos (q+k_1)-\cos k_2\big)(t-\tau)} \mathcal{G}(k_1, k_1, q) \\
&+e^{-\frac{i2J}{\hbar}\big(\cos k_1-\cos (q+k_2) \big)(t-\tau)} \mathcal{G}(k_2, k_2, q)\bigg\} ,
\end{split} 
\end{equation}
which represents the $k$-local memory effect where the propagation of the density matrix interacts only with the local density matrix in reciprocal space, $\langle \tilde{\rho}(k_1,k_2;\tau) \rangle_\mathrm{E}$ while the disorder-induced interactions are encoded in the memory kernel.
On the other hand, $\mathcal{K}_\mathrm{scat}$ is defined as 
\begin{align}
\nonumber
& \mathcal{K}_\mathrm{scat}(k_1,k_2,q;t,\tau)= \frac{1}{2\pi\hbar^2}\mathcal{G}(k_1, k_2, q)\\
&\times  \bigg\{e^{-\frac{i2J}{\hbar}\big(\cos (q+k_1)-\cos k_2\big)(t-\tau)} +e^{-\frac{i2J}{\hbar}\big(\cos k_1-\cos (q+k_2) \big)(t-\tau)}\bigg\} ,
\end{align}
which represents $k$-nonlocal memory effect stemming from momentum scattering in reciprocal space. Namely, the propagation of the density matrix involves disorder-induced interactions with the entire Brillouin zone.

\subsection{Laplace transform of the reciprocal-space LvN equation}
\label{Sec:Laplace_Dyn_Eq}
To investigate the time evolution of $\langle \tilde{\rho}(k_1,k_2;t) \rangle_\mathrm{E}$, we apply the Laplace transform to Eq.~(\ref{Eq:Integro-Differ}), defined as $\mathcal{L}[ \tilde{f}(t)] = \tilde{f}(p) = \int_0^\infty e^{-pt} \tilde{f}(t) dt$. By introducing the variable substitution $u = k_1 - k_2$ for the relative momentum and $s = (k_1 + k_2)/2$ for the center-of-mass momentum, we define the Laplace transform of the ensemble-averaged density matrix by $F(u,s;p) \equiv \mathcal{L}[\langle \tilde{\rho}(s+\frac{u}{2}, s-\frac{u}{2}; t) \rangle_\mathrm{E}]$ and convert Eq.~(\ref{Eq:Integro-Differ}) into an algebraic equation in the $p$-domain (see Appendix~\ref{Appendix:Laplace_Tran_Integro-Differ} for the details):
\begin{align}
\nonumber
    p F(u,s;p) = &
    \langle \tilde{\rho}(k_1,k_2;t=0) \rangle_\mathrm{E}  + \frac{i4J}{\hbar} \sin (s) \sin\left(\frac{u}{2}\right) F(u,s;p) \\
    \nonumber & +{K}_\mathrm{loc}(u,s;p)  F(u,s;p)  \\  
    & + \int_{-\pi}^{\pi} dq{K}_\mathrm{scat}(u,s,q;p) F(u,s+q;p) .
\label{Eq:Laplace_Integro-Differ_us}
\end{align}
The corresponding memory kernels in the $p$-domain can be expressed as 
\begin{align}
\nonumber
    &{K}_\mathrm{loc}(u,s;p)\\
    &=-\frac{1}{2\pi\hbar^2}\sum_{\sigma=\pm}\int_{-\pi}^{\pi} dq {K}_{\sigma}(u,s,q;p){G}(0, s\pm u/2, q) ,
\end{align}
and 
\begin{equation}
    {K}_\mathrm{scat}(u,s,q;p)=\frac{1}{2\pi\hbar^2}{G}(u, s, q)\sum_{\sigma=\pm}{K}_{\sigma}(u,s,q;p) ,
\end{equation}
where we have ${G}(u,s,q)\equiv\mathcal{G}(s+\frac{u}{2},s-\frac{u}{2},q)=g_0+2g_1[\cos(q+2s)+\cos(u)]$ as shown in Eq.~(\ref{Eq:Reciprocal-space disorder_kernel_us})  
and ${K}_{\sigma}(u,s,q;p)$ is defined as follows:
\begin{align}
\nonumber
{K}_{\sigma}(u,s,q;p)
&=\frac{1}{
p-\frac{i4J}{\hbar}
\sin\!\left(\frac{q+2s}{2}\right)
\sin\!\left(\frac{\sigma q+u}{2}\right)
}\\
&=\frac{1}{p}
\sum_{n=0}^{\infty}
\left[
\frac{i4J}{p\hbar}
\sin\!\left(\frac{q+2s}{2}\right)
\sin\!\left(\frac{\sigma q+u}{2}\right)
\right]^n,
\label{Eq:R_sigma_series}
\end{align}
For the second line in Eq.~\eqref{Eq:R_sigma_series}, we take a geometric series expansion with the convergence condition $p > \frac{4J}{\hbar}$ (i.e., $t<\frac{\hbar}{4J}$). Since we are interested in short-time exciton dynamics, we consider the large-$p$ limit of the Laplace transform to determine the asymptotic behavior of $\langle \tilde{\rho}(k_1,k_2;t) \rangle_\mathrm{E}$ as $t \to 0$. Consequently, we truncate Eq.~(\ref{Eq:R_sigma_series}) to the leading order, i.e. ${K}_{\sigma}(u,s,q;p) \approx 1/p$, corresponding to the so-called short-time approximation\cite{cui2023short}. This truncation allows Eq.~(\ref{Eq:Laplace_Integro-Differ_us}) to be approximated as:
\begin{align}
\nonumber
    p F(u,s;p) =& 
    \langle \tilde{\rho}(k_1,k_2;t=0) \rangle_\mathrm{E}  + \frac{i4J}{\hbar} \sin (s) \sin\left(\frac{u}{2}\right) F(u,s;p) \\
    \nonumber & - \frac{2 (g_0+2g_1)}{ \hbar^2 p} F(u,s;p)  \\
    &+ \frac{1}{ \pi\hbar^2 p} \int_{-\pi}^{\pi} dq  {G}(u, s, q) F(u,s+q;p) .
\label{Eq:Laplace_Integro-Differ_us_Approx}
\end{align}

To further simplify Eq.~(\ref{Eq:Laplace_Integro-Differ_us_Approx}), we introduce the bare propagator that includes the $p$-local memory term: 
\begin{equation}
D(u,s,p) =  \left[ p + \frac{2(g_0+2g_1)}{ \hbar^2 p}  -  \frac{i4J}{\hbar} \sin(s) \sin(\frac{u}{2}) \right]^{-1}.
\end{equation} 
By invoking the translational invariance of the integral (specifically, the shift $q\rightarrow q-s$), we have the relation $\int_{-\pi}^{\pi} dq\, {G}(u, s, q) F(u,s+q;p) = \int_{-\pi}^{\pi} dq \, {G}(u, s, q-s) F(u,q;p)$ and Eq.~\eqref{Eq:Laplace_Integro-Differ_us_Approx} becomes:
\begin{align}
\nonumber
    & F(u,s;p) = 
    D(u,s,p) \langle \tilde{\rho}(k_1,k_2;t=0) \rangle_\mathrm{E}  \\
\nonumber
    &  + \frac{D(u,s,p)}{ \pi\hbar^2 p}  \left[g_0 + 2g_1 \cos(u) \right] \int_{-\pi}^{\pi} dq  F(u,q;p) \\
    &  + \frac{2g_1 D(u,s,p)}{ \pi\hbar^2 p} \int_{-\pi}^{\pi} dq   \left[   \cos(s)\cos(q) -  \sin(s)\sin(q) \right] F(u,q;p) .
\label{Eq:Laplace_Integro-Differ_us_Approx_Solvable}
\end{align}

\subsection{Solving the reciprocal-space LvN equation}
To solve Eq.~(\ref{Eq:Laplace_Integro-Differ_us_Approx_Solvable}), we define the following three integral moments\cite{madhukar1977exact}:
\begin{align}
\label{Eq:Define_Omega_1}
    &\Omega_1(u;p) = \hat{O}_1 F(u, q ; p) = \frac{1}{2\pi} \int_{-\pi}^{\pi} dq F(u, q ; p) ,  \\
    &\Omega_2(u;p)  =  \hat{O}_2 F(u, q ; p) = \frac{1}{2\pi} \int_{-\pi}^{\pi} dq \cos(q) F(u, q ; p),  \\
    &\Omega_3(u;p)  = \hat{O}_3 F(u, q ; p) = \frac{1}{2\pi} \int_{-\pi}^{\pi} dq \sin(q) F(u, q ; p).
\label{Eq:Define_Omega_3}
\end{align}
Substituting these into Eq.~(\ref{Eq:Laplace_Integro-Differ_us_Approx_Solvable}), the equation for $F(u,s;p)$ becomes:
\begin{align}
\nonumber 
    & F(u,s;p) = 
    D(u,s,p) \langle \tilde{\rho}(k_1,k_2;t=0) \rangle_\mathrm{E}  \\
\nonumber
    & \qquad + \frac{2D(u,s,p)}{ \hbar^2 p}  \left[g_0 + 2g_1 \cos(u) \right] \Omega_1(u;p) \\
    & \qquad + \frac{4g_1 D(u,s,p)}{ \hbar^2 p}  \left[   \cos(s)\Omega_2(u;p) -  \sin(s)\Omega_3(u;p) \right] .
\label{Eq:Laplace_Integro-Differ_us_Approx_Solvable_Omega}
\end{align}

Equation~(\ref{Eq:Laplace_Integro-Differ_us_Approx_Solvable_Omega}) can be broken down into a set of linear equations by applying the integral operators $\hat{O}_j$, as defined in Eqs.~(\ref{Eq:Define_Omega_1})--(\ref{Eq:Define_Omega_3}), to both sides of the equation. By treating $s$ as the integration variable (e.g., $\Omega_1(u;p) = \frac{1}{2\pi} \int_{-\pi}^{\pi} ds F(u, s ; p)$), we obtain a closed system of three linear algebraic equations for the variables $\Omega_1 \equiv \Omega_1(u;p)$, $\Omega_2 \equiv \Omega_2(u;p)$, and $\Omega_3 \equiv \Omega_3(u;p)$:
\begin{equation}
\begin{bmatrix}
1 - \frac{\Gamma_0+\Gamma_1\cos u}{p} C_1 & -\frac{\Gamma_1}{p}C_2 & \frac{\Gamma_1}{p}C_3 \\
-\frac{\Gamma_0+\Gamma_1\cos u}{p} C_2 & 1 - \frac{\Gamma_1}{p}C_4 & \frac{\Gamma_1}{p}C_5 \\
-\frac{\Gamma_0+\Gamma_1\cos u}{p} C_3  & - \frac{\Gamma_1}{p}C_5 & 1 +  \frac{\Gamma_1}{p}C_6
\end{bmatrix}
\begin{bmatrix}
\Omega_1 \\
\Omega_2 \\
\Omega_3
\end{bmatrix}
=
\begin{bmatrix}
S_1  \\
S_2 \\
S_3 
\end{bmatrix},
\label{Eq:System_Algebraic}
\end{equation}
where we define the parameters $\Gamma_0 = 2g_0/\hbar^2$ and $\Gamma_1 = 4g_1/\hbar^2$. The source terms $S_j$ related to the initial condition of $ \hat{\rho}(t=0) $ are evaluated as follows: 
\begin{align}
\nonumber
    &S_1 = \frac{1}{2\pi} \int_{-\pi}^{\pi} ds D(u, s; p)\langle \tilde{\rho}(k_1,k_2;t=0) \rangle_\mathrm{E} , \\
\nonumber
    &S_2 = \frac{1}{2\pi} \int_{-\pi}^{\pi} ds \cos(s) D(u, s; p)\langle \tilde{\rho}(k_1,k_2;t=0) \rangle_\mathrm{E} , \\
    &S_3  = \frac{1}{2\pi} \int_{-\pi}^{\pi} ds \sin(s) D(u, s; p)\langle \tilde{\rho}(k_1,k_2;t=0) \rangle_\mathrm{E} .
    \label{Eq:Source}
\end{align}
The coefficients $C_j$ are evaluated as follows: 
\begin{align}
\nonumber
    &C_1 = \frac{1}{2\pi} \int_{-\pi}^{\pi} ds D(u, s; p) = \frac{1}{\sqrt{A^2-B^2}}, \\
\nonumber
    &C_2 = \frac{1}{2\pi} \int_{-\pi}^{\pi} ds \cos(s) D(u, s; p) = 0, \\
\nonumber
    &C_3 =\frac{1}{2\pi} \int_{-\pi}^{\pi} ds \sin(s) D(u, s; p) = \frac{1}{B} \left( 1- \frac{A}{\sqrt{A^2-B^2}} \right), \\
\nonumber
    &C_4 =\frac{1}{2\pi} \int_{-\pi}^{\pi} ds \cos^2(s) D(u, s; p) = \frac{1}{B^2} \left( A- {\sqrt{A^2-B^2}} \right), \\
\nonumber
    &C_5 = \frac{1}{2\pi} \int_{-\pi}^{\pi} ds \cos(s) \sin(s) D(u, s; p) = 0, \\
    &C_6 = \frac{1}{2\pi} \int_{-\pi}^{\pi} ds \sin^2(s) D(u, s; p) = -\frac{  A}{B^2} \left( 1- \frac{A}{\sqrt{A^2-B^2}} \right) ,
\label{Eq:Coefficients}
\end{align}
where
\begin{align}
\nonumber
    &A \equiv A(p) = p + \frac{\Gamma_0+ \Gamma_1}{p} , \\ 
    &B \equiv B(u) = -i \frac{4J}{\hbar} \sin(\frac{u}{2}) .
\label{Eq:Coeff_AB}
\end{align}

Given that $C_2 = C_5 = 0$, the equation for $\Omega_2$ decouples from the system in Eq.~(\ref{Eq:System_Algebraic}). Furthermore, as the physical observables $\langle x \rangle$ and $\langle x^2 \rangle$ are directly determined by $\Omega_1$, we employ Cramer’s rule to obtain the analytical solution for $\Omega_1$:
\begin{equation}
\Omega_1(u;p) = \frac{1}{\Delta} \mathrm{det}
\begin{pmatrix}
S_1 & \frac{\Gamma_1}{p}C_3 \\
S_3  & 1 +  \frac{\Gamma_1}{p}C_6
\end{pmatrix}  ,
\label{Eq:Omega_1_Formal}
\end{equation}
where
\begin{equation}
\Delta = \mathrm{det}
\begin{pmatrix}
1 - \frac{\Gamma_0+\Gamma_1\cos u}{p} C_1 & \frac{\Gamma_1}{p}C_3 \\
-\frac{\Gamma_0+\Gamma_1\cos u}{p} C_3  & 1 +  \frac{\Gamma_1}{p}C_6 
\end{pmatrix} .
\label{Eq:Determinant}
\end{equation}

\subsection{Short-time behavior of $\langle x \rangle$ and $\langle x^2 \rangle$ under two specific initial conditions}
\label{Sec:Short-time_behavior_Observables}
As shown in Appendix~\ref{Appendix:Site_representation_Omega_1}, $\Omega_1(u,p)$ corresponds to the Fourier transform of site populations, i.e., $\Omega_1(u,p) = \sum_{l}  e^{- i u l} \langle {\rho}_{l,l}(p) \rangle_\mathrm{E}$. Consequently, the average position $\langle x(p) \rangle$ and the second spatial moment $\langle x^2(p) \rangle$ in the $p$-domain can be obtained by evaluating the first and second derivatives of the generating function with respect to the reciprocal variable $u$ at the origin:
\begin{align}
    \langle x^m(p) \rangle = (ia)^m \left. \frac{\partial^m \Omega_1(u;p)}{\partial u^m} \right|_{u=0} = \sum_{l} (al)^m \langle {\rho}_{l,l}(p)  \rangle_\mathrm{E}.
\label{Eq:Spatial_Moment_Def}
\end{align}

To obtain an explicit expression for $\Omega_1$, the initial condition $\langle \tilde{\rho}(k_1,k_2;t=0) \rangle_\mathrm{E} = \sum_{l,r} e^{-ik_1 l + i k_2 r} \rho_{l,r}(0)$ must be specified. In this study, we consider two representative initial conditions: (\textit{i}) a localized excitation and (\textit{ii}) a moving Gaussian wavepacket. 

\subsubsection{Local excitation}
\label{Sec:Local_Excitation}
We first consider the case of a strictly localized initial population at the origin, i.e., $\rho_{l,r}(t=0) = \delta_{l,0}\delta_{r,0}$. In this limit, the initial reciprocal-space density reduces to unity, $\langle \tilde{\rho}(k_1,k_2;0) \rangle_\mathrm{E} = 1$, implying that all momentum states across the Brillouin zone are equally weighted. Applying this condition to the coefficients $S_1$ and $S_3$ defined in Eq.~(\ref{Eq:Source}), we find that they reduce to $S_1 = C_1$ and $S_3=C_3$. Consequently, the explicit expression for $\Omega_1$ becomes (see Appendix~\ref{Appendix:Omega1_Local_Excit} for the details):
\begin{align}
\Omega_1(u;p)=\frac{X}{XY-B^2},
\label{Eq:Omega1_XY_form}
\end{align}
with
\begin{align}
X=A+\sqrt{A^2-B^2}+\frac{\Gamma_1}{p},
\qquad
Y=A-\frac{\Gamma_0+\Gamma_1\cos u}{p}.
\label{Eq:XY_explicit}
\end{align}

The parity of $\Omega_1(u;p)$ is determined by the $u$-dependence of its components: $A$ is independent of $u$, while $B^2 \propto \sin^2(u/2)$ and $Y$ (a function of $\cos u$) are both even. Consequently, $\Omega_1(u;p) = \Omega_1(-u;p)$ is an even function of $u$. Due to this symmetry, the first derivative vanishes at the origin ($\Omega'_1(0;p) = -\Omega'_1(0;p) = 0$), resulting in a vanishing average position, $\langle x \rangle = 0$, for the locally excited initial condition.

To evaluate the second derivative of $\Omega_1(u;p)$ at $u=0$, we first expand the auxiliary functions $B$, $X$, and $Y$ [Eqs.~(\ref{Eq:Coeff_AB}) and (\ref{Eq:XY_explicit})] to the second order. Following the application of the quotient rule and subsequent simplification (detailed in Appendix~\ref{Appendix_Second_Derive_Omega1_Local}), we arrive at:
\begin{align}
\left. \frac{\partial^2 \Omega_1(u,p)}{\partial u^2 } \right|_{u=0} & =  -\frac{\Gamma_1}{p^3} - \frac{8J^2}{p \left( 2p^2 + 2\Gamma_0 + 3\Gamma_1 \right) \hbar^2} .
\label{Eq:2nd_Derive_Omega_1}
\end{align}
Subsequently, the second spatial moment in the time domain $\langle x^2(t) \rangle$ is determined via the inverse Laplace transform of Eq.~(\ref{Eq:2nd_Derive_Omega_1}):
\begin{align}
\nonumber
    &\langle x^2(t) \rangle = \mathcal{L}^{-1} \left\{ (ia)^2 \left.\frac{\partial^2 \Omega_1(u,p)}{\partial u^2 } \right|_{u=0} \right\} \\
    & = a^2 \left\{ \frac{\Gamma_1}{2}t^2 + \frac{16 J^2}{\hbar^2(2\Gamma_0+3\Gamma_1)} \left[ \sin^2\left(\sqrt{\frac{2\Gamma_0+3\Gamma_1}{8}}\,t\right)
\right] \right\} .
\label{Eq:x_square_NotSimple}
\end{align}
In the short-time limit, $t \ll \sqrt{8 / (2\Gamma_0 + 3\Gamma_1)}$, a Taylor expansion of Eq.~(\ref{Eq:x_square_NotSimple}) yields the simplified final expression:
\begin{align}
    \langle x^2(t) \rangle & \approx \, a^2 \left\{ \frac{\Gamma_1}{2}t^2 + \frac{2 J^2}{\hbar^2 } t^2 \right\}  = \frac{2(g_1 + J^2)a^2 }{\hbar^2}  t^2.
\label{Eq:x_square_Simple}
\end{align}

Equation~(\ref{Eq:x_square_Simple}) is the first key result of our analysis, showing that the initial delocalization is purely ballistic, $\langle x^2(t) \rangle \propto t^2$, and is governed jointly by the deterministic coupling $J$ and the off-diagonal disorder strength $g_1$. Interestingly, the absence of diagonal disorder $g_0$ from this leading-order result indicates that on-site energy fluctuations do not affect early-time spreading. Off-diagonal disorder thus leads to enhancing the initial coherent expansion before the diffusive regime emerges.

\subsubsection{Moving Gaussian wavepacket}
Next, we consider a fully coherent Gaussian wavepacket characterized by a finite spatial width $w_0$ and a dimensionless parallel momentum $k_\parallel$. Unlike the localized excitation discussed previously, this initial condition features non-vanishing off-diagonal elements, thereby encoding both phase coherence and a preferred propagation direction from the outset. The corresponding initial real-space density matrix is defined as:
\begin{align}
\rho_{l,r}(0)
=
\mathcal{N}^2
\exp\!\left[-\frac{l^2+r^2}{2w_0^2}\right]
e^{ik_\parallel(l-r)},
\label{Eq:Gaussian_initial_density_matrix}
\end{align}
where $\mathcal{N}$ denotes the normalization constant. 

In this study, we assume a wavepacket broad enough (i.e., $w_0\gg 1$) that its discrete Fourier transform is well-approximated by the continuum limit. As detailed in Appendix~\ref{Appendix:Gaussian_initial_condition}, the corresponding density in reciprocal space takes the form:
\begin{align}
\tilde{\rho}(u,s;0)
=
2 \sqrt{\pi} w_0 \exp\!\left[-\frac{w_0^2}{4}u^2\right] \exp\!\left[-w_0^2(s-k_\parallel)^2\right].
\label{Eq:Gaussian_initial_us_main}
\end{align}

Equation~(\ref{Eq:Gaussian_initial_us_main}) illustrates that, in contrast to the localized excitation treated in Sec.~\ref{Sec:Local_Excitation}, the initial state of a Gaussian wavepacket selectively populates a narrow momentum window centered at $k_\parallel$, with a momentum spread inversely proportional to the spatial width $w_0$.

Substituting this Gaussian initial condition, Eq.~(\ref{Eq:Gaussian_initial_us_main}), into the source coefficients of Eq.~(\ref{Eq:Source}) yields the necessary inputs for evaluating the integral moment $\Omega_1(u;p)$ defined in Eq.~(\ref{Eq:Omega_1_Formal}). Following the derivations outlined in Appendix~\ref{Appendix:Omega1_Gaussian}, we arrive at:
\begin{align}
    \Omega_1(u;p) = e^{-\frac{w_0^2}{4}u^2}  \frac{ AX + \frac{\Gamma_1}{p}B\sin{k_\parallel} -B^2   }{ (A+B\sin{k_\parallel})(XY-B^2)  } .
\label{Eq:Omega1_Gaussian_Final}
\end{align}
Distinct from the localized scenario, Eq.~(\ref{Eq:Omega1_Gaussian_Final}) incorporates an asymmetric term $B\sin k_\parallel \sim \sin(u/2)k_\parallel$ in both its numerator and denominator. Consequently, $\Omega_1(u;p)$ lacks invariance under the parity transformation $u\to -u$ whenever $k_\parallel\neq 0$. This explicit breaking of parity symmetry is the underlying mechanism responsible for the non-zero first spatial moment $\langle x \rangle$.

By performing a Taylor expansion of Eq.~(\ref{Eq:Omega1_Gaussian_Final}) around $u=0$ and applying the quotient rule (see Appendix~\ref{Appendix:Gaussian_small_u_expansion} for the details), we extract the first and second spatial moments:
\begin{align}
\langle x(p)\rangle
&= \frac{-4aJ \sin k_\parallel }{ \hbar ( 2p^2 +2\Gamma_0 +3\Gamma_1 ) },
\label{Eq:xp_Gaussian_def}
\\
\langle x^2(p)\rangle
&= a^2\left(
\frac{  w_0^2}{2p}
+ \frac{8J^2 \sin^2k_\parallel}{ \hbar^2 p^3 } +\frac{\Gamma_1}{p^3}  \right).
\label{Eq:x2p_Gaussian_def}
\end{align}
Then we apply an inverse Laplace transform, which yields the corresponding short-time dynamical behavior:
\begin{align}
\langle x(t)\rangle
&= - a \frac{2 J \sin k_\parallel}{\hbar} t,
\label{Eq:x_t_Gaussian_placeholder}
\\
\langle x^2(t)\rangle
&= a^2 \left[ \frac{w_0^2}{2} + \left( \frac{4 J^2 \sin^2 k_\parallel}{\hbar^2} + \frac{ 2g_1 }{\hbar^2 } \right) t^2    \right].
\label{Eq:x2_t_Gaussian_placeholder}
\end{align}

Equations~(\ref{Eq:x_t_Gaussian_placeholder}) and \eqref{Eq:x2_t_Gaussian_placeholder} are the second key results of our analysis. Eq.~\eqref{Eq:x_t_Gaussian_placeholder} reveals that the non-zero dimensionless parallel momentum $k_\parallel$ induces a ballistic drift of the wavepacket's center of mass $\langle x(t)\rangle$. Concurrently, Eq.~(\ref{Eq:x2_t_Gaussian_placeholder}) demonstrates that the second spatial moment $\langle x^2(t)\rangle$ decomposes naturally into an initial Gaussian wavepacket's width and a time-dependent $t^2$ expansion. Within the dynamical $t^2$ term, the spatial broadening of the wavepacket over time is driven by two terms: the effect of translation ($J^2\sin^2k_\parallel$) and the off-diagonal disorder ($g_1$).
Note that the diagonal disorder $g_0$ is also absent in this leading-order result, suggesting a consistent result that the short-time spreading of a moving Gaussian wavepacket is not sensitive to on-site energy fluctuations.

\section{Numerical Demonstration and Discussion} \label{Sec:Discussion}

In this section, we numerically validate the analytical expressions derived in Eqs. (\ref{Eq:x_square_Simple}), (\ref{Eq:x_t_Gaussian_placeholder}), and (\ref{Eq:x2_t_Gaussian_placeholder}). Numerical simulations are performed with a grid of $N_{\mathrm{lattice}} = 4000$ points and randomly sampled $\beta_{mn}$ following the disorder correlation, Eq.~\eqref{Eq:Covariance_beta}.
All physical observables are averaged over 4000 independent trajectories to ensure statistical convergence.

\subsection{Local excitation initial condition}
\label{Sec:Numerical_Local_Exc}

\subsubsection{Short-time ballistic spread and insensitivity to diagonal disorder}
To numerically validate the insensitivity of $\langle x^2(t) \rangle$ on $g_0$, we consider a molecular aggregate with a deterministic nearest-neighbor coupling of $J = -10$~meV [Eq.~(\ref{Eq:NN_Coupling_J})], typical for dipole-dipole interactions in J-aggregates, and set the off-diagonal disorder strength to $g_1 = (10~\text{meV})^2$. The lattice constant is chosen to be $a = 1$~nm for convenience.

\begin{figure}
   \includegraphics[width=1.0\linewidth]{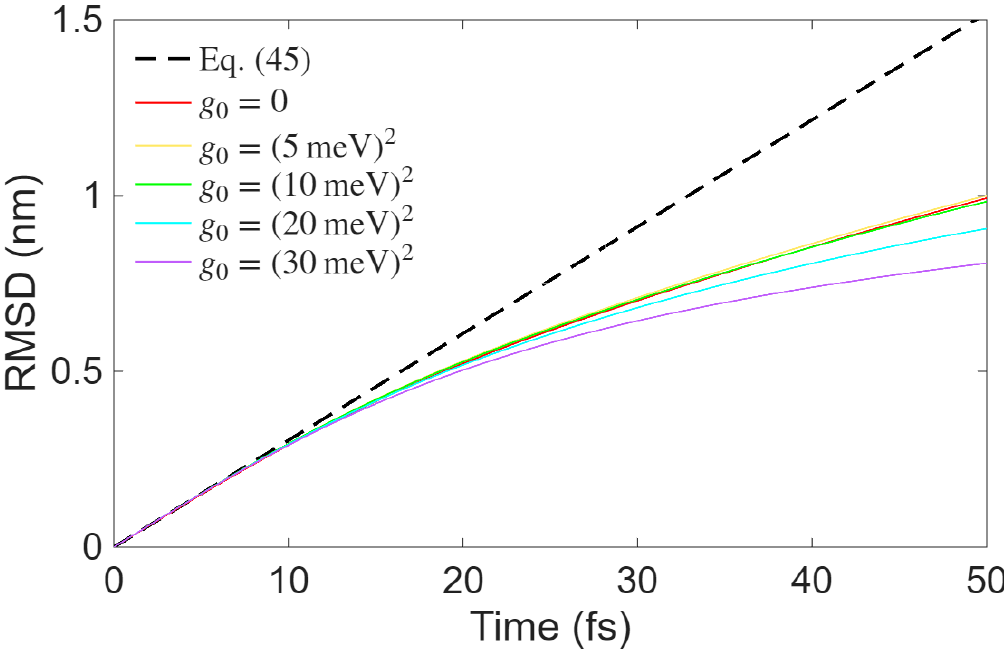}
    \caption{Time evolution of the RMSD with the local excitation initial condition for various diagonal disorder $g_0$, while maintaining a constant coupling $J$ and off-diagonal disorder $g_1$. The universal overlap of all cases for $t \lesssim 10 \,\text{fs}$ validates the short-time analytical formula Eq.~(\ref{Eq:RMSD_linear}). The progressive deviation of the RMSD from the linear limit highlights that increasing the diagonal disorder magnitude $g_0$ narrows the temporal range of validity for the ballistic approximation. } \label{Fig:Local_Exc_g0_Inv} 
\end{figure}

Because Eq.~(\ref{Eq:x_square_Simple}) implies a ballistic regime where $\langle x^2(t) \rangle \propto t^2$, we characterize the delocalization using the root mean square displacement (RMSD), defined as $\text{RMSD}(t) = \sqrt{\langle x^2(t) \rangle - \langle x(t) \rangle^2}$. For a locally excited initial condition ($\langle x(t) \rangle = 0$), the analytical short-time RMSD is linear in $t$:
\begin{align}
\mathrm{RMSD}(t) = a \frac{\sqrt{2(J^2+g_1)}}{\hbar} t,
\label{Eq:RMSD_linear}
\end{align}
In Fig.~\ref{Fig:Local_Exc_g0_Inv}, we present the numerical evolution of the RMSD for on-site disorder strengths $g_0$ ranging from $0$ to $(30~\text{meV})^2$. The simulations show excellent agreement with the analytical prediction (black dashed line) for approximately the first $\sim 10$~fs. Remarkably, even when $g_0$ exceeds $g_1$ by almost an order of magnitude, all trajectories still collapse onto the same linear curve at early times. This confirms that diagonal disorder does not influence the initial ballistic spread. Beyond this regime, the curves diverge as the specific magnitude of $g_0$ begins to dictate the transition toward diffusive transport.

\subsubsection{Equivalent contribution of $J^2$ and $g_1$ in the short-time limit}

\begin{figure}
   \includegraphics[width=0.95\linewidth]{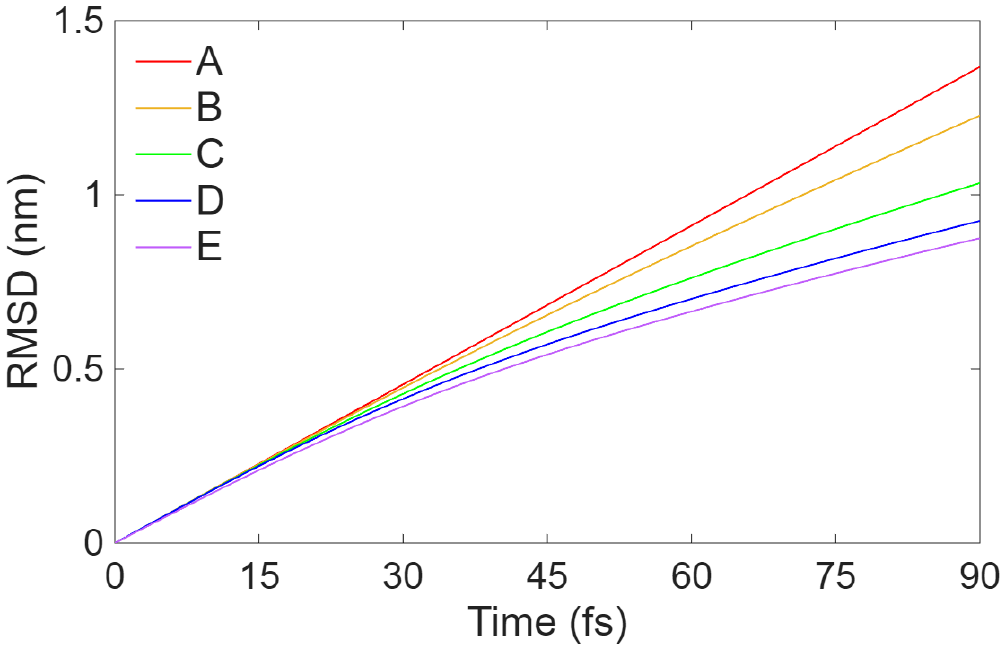}
    \begin{ruledtabular}
    \begin{tabular}{ccc}
    Case Index & $J$ (meV) & $g_1$ ($\mathrm{meV^2}$) \\
    \hline
    A & -7.07 & 0 \\
    B & -6.75 & $(2.11)^2$\\
    C & -6.00 & $(3.74)^2$\\
    D & -5.00 & $(5.00)^2$\\
    E & -4.44 & $(5.50)^2$\\
    \end{tabular}
    \end{ruledtabular}
    \caption{Time evolution of the RMSD with the local excitation initial condition for several $(J,g_1)$ combinations with fixed effective coupling $\sqrt{J^2+g_1}\approx 7.07\,\text{meV}$. The table shows the combinations ranging from a purely ordered lattice (Case A, $g_1=0$) to a disorder-dominated lattice (Case E, $g_1> J^2$). The convergence of all curves at early times confirms the analytical prediction of Eq.~(\ref{Eq:RMSD_linear}) and shows that the initial expansion depends only on the total effective coupling strength. The subsequent departure from the common linear behavior indicates that, although $J$ and $g_1$ initially play an equivalent role, increasing off-diagonal disorder progressively shortens the ballistic regime. } \label{Fig:Local_Exc_J_g1_Equiv}
\end{figure}

To further investigate the early-time transport behavior, we consider a series of $(J, g_1)$ parameter sets with a fixed effective coupling strength $\sqrt{J^2 + g_1} \approx 7.07 \text{ meV}$ and zero diagonal disorder $g_0=0$. The parameter set ranges from the ballistic transport regime (Case A: $g_1=0$) to the disorder-dominated regime (Case E: $g_1 > J$). At short times ($t \lesssim 15 \text{ fs}$), all the cases in Fig.~\ref{Fig:Local_Exc_J_g1_Equiv} converge to the same linear curve. This convergence indicates that, at the onset of expansion, the exciton is effectively insensitive to the specific microscopic transport mechanism. Whether the off-diagonal contribution arises from deterministic coherent coupling $J$ or random fluctuations $g_1$, the resulting RMSD is determined solely by the magnitude of the total effective coupling.

As time evolves beyond this initial regime, however, the RMSD progressively deviates from the common linear limit (Case A). This separation shows that although $J$ and $g_1$ play equivalent roles in driving the initial expansion, increasing the off-diagonal disorder substantially reduces the time window over which the ballistic approximation remains valid. Physically, this window marks the stage at which cumulative phase-breaking scattering events begin to resolve the underlying disorder, thereby shortening the exciton mean free path and truncating the ballistic regime. By comparing the sustained ballistic expansion of the coherent benchmark (Case A)\cite{merrifield1958propagation}, $\text{RMSD} = \frac{\sqrt{2J^2}}{\hbar} t$, with the much earlier diffusive behavior observed in Case E, we clearly illustrate how off-diagonal fluctuations induce the crossover from ballistic to diffusive transport.


\subsection{Moving Gaussian wavepacket initial condition}
\label{Sec:Numerical_Guassian_Exc}

\subsubsection{Short-time behavior of wavepacket's center of mass $\langle x(t) \rangle$}

To verify the analytical prediction for the first spatial moment, we investigate the short-time dynamics of a Gaussian wavepacket with an initial width of $w_0=40$, chosen to ensure the validity of the continuum approximation. The simulation parameters are set to $J=-10~\mathrm{meV}$ and $g_0=g_1=(10~\mathrm{meV})^2$, corresponding to a regime in which the deterministic coupling and random fluctuations are of comparable magnitude. 

\begin{figure}
   \includegraphics[width=0.99\linewidth]{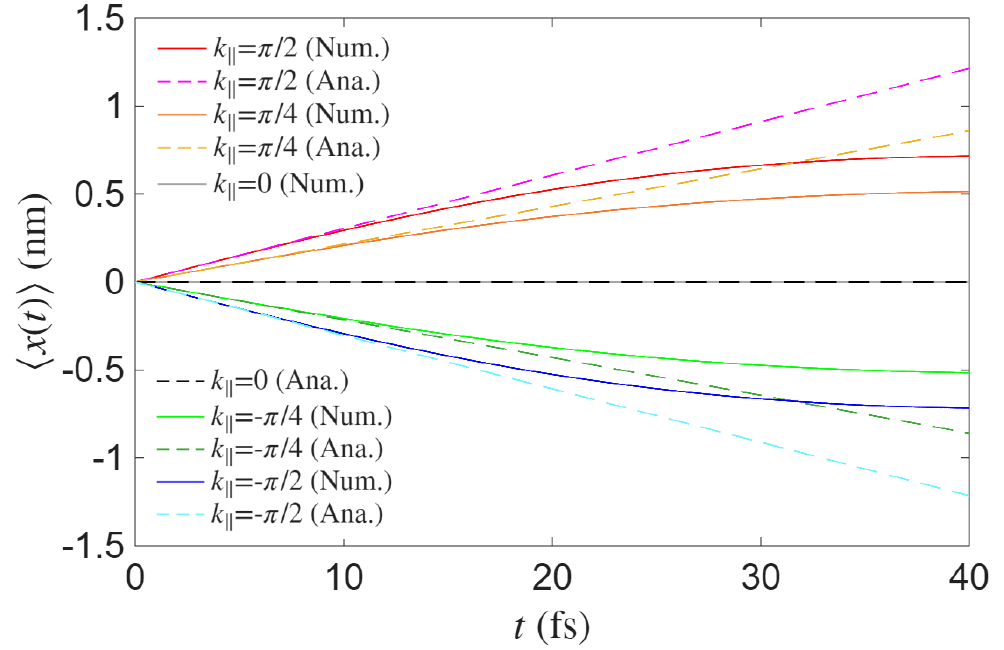}
    \caption{Time evolution of the first spatial moment $\langle x(t)\rangle$ for several values of the dimensionless parallel momentum $k_\parallel \in [-\pi/2,\pi/2]$ of the initial Gaussian wavepacket. The rest parameters are fixed at $J=-10~\mathrm{meV}$, $g_0=g_1=(10~\mathrm{meV})^2$, and $w_0=40$. Solid lines are numerical results; dashed lines are analytical predictions. The close agreement between the two confirms that the short-time behavior of $\langle x\rangle$ is governed solely by the coherent coupling $J$ and is insensitive to the disorder strengths $g_0$ and $g_1$. } \label{Fig:Gaussian_Exc_x_ave}
\end{figure}

Figure~\ref{Fig:Gaussian_Exc_x_ave} shows $\langle x(t) \rangle$ for various dimensionless parallel momentum $k_\parallel$ values.
As predicted by Eq.~(\ref{Eq:x_t_Gaussian_placeholder}), the center of mass exhibits a linear time dependence in the short-time regime ($t\lesssim 15~\mathrm{fs}$), with a group velocity given by $v_{\mathrm g} = d\langle x(t) \rangle/dt \propto -2J\sin k_\parallel$. This trend is clearly reflected in Fig.~\ref{Fig:Gaussian_Exc_x_ave} where the positive and negative values of $k_\parallel$ lead to drift in opposite directions, while the $k_\parallel=0$ case remains centered around $\langle x\rangle=0$. The numerical results (solid lines) closely match analytical predictions (dashed lines), confirming that early-time drift is governed solely by the coherent coupling $J$.

The significance emerges upon contrasting with the second moment $\langle x^2(t)\rangle$ in Eq.~(\ref{Eq:x2_t_Gaussian_placeholder}). While the first moment $\langle x(t)\rangle$ is independent of $g_0$ and $g_1$, $\langle x^2(t)\rangle$ includes both coherent transport ($4J^2\sin^2 k_\parallel$) and off-diagonal disorder ($2g_1$). Hence, off-diagonal fluctuations only symmetrically broaden the exciton distribution without affecting the group velocity. Notably, since the RMSD exclude the coherent part of $\langle x^2(t)\rangle$, the initial increase in $\text{RMSD}=\sqrt{\langle x^2(t)\rangle-\langle x(t)\rangle^2}$ directly probes the off-diagonal disorder strength $g_1$, independent of $J$ and $g_0$.

\subsubsection{Short-time RMSD: Insensitivity to diagonal disorder}
The Gaussian wavepacket described in Eq.~(\ref{Eq:Gaussian_initial_density_matrix}) can be generated using a pulsed Gaussian laser beam with an angular wavenumber $\bar{k} = \sqrt{\bar{k}_\parallel^2+\bar{k}_\perp^2} = 2\pi/\lambda$. For a typical blue-light wavelength ($\lambda = 400$ nm) and a lattice constant ($a = 1$ nm), the dimensionless parallel momentum satisfies $k_\parallel = a\bar{k}_\parallel \leq 2\pi a / \lambda \approx 0.016$. Given that this value is negligible, we set $k_\parallel = 0$ hereafter for simplicity. According to Eqs. (\ref{Eq:x_t_Gaussian_placeholder}) and (\ref{Eq:x2_t_Gaussian_placeholder}), the analytical short-time $\text{RMSD}(t) = \sqrt{ \langle x^2(t) \rangle - \langle x(t) \rangle^2 }$ evolves quadratically with $t$:
\begin{align}
 \text{RMSD}(t) = a \sqrt{ \frac{w_0^2}{2} +  \frac{ 2g_1 }{\hbar^2 }  t^2  } \approx a \left[ \sqrt{\frac{w_0^2}{2} } +  \frac{ \sqrt{2} g_1 }{ w_0 \hbar^2 }  t^2 \right]   
\label{Eq:RMSD_Gaussian}
\end{align}

\begin{figure}
   \includegraphics[width=0.98\linewidth]{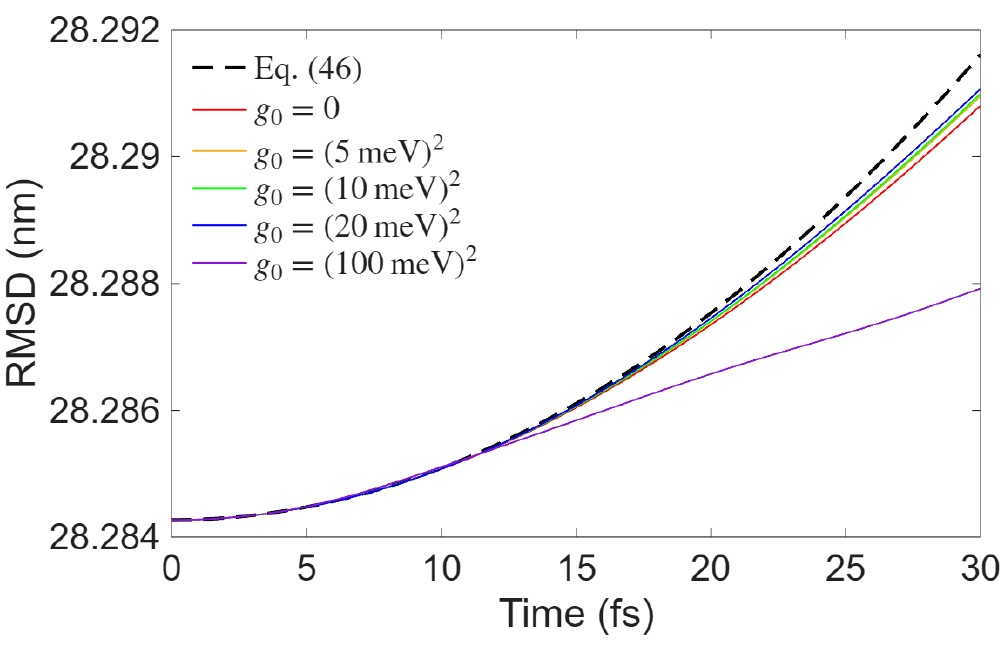}
    \caption{ Time evolution of the RMSD with the Gaussian wavepacket initial condition for various on-site disorder strengths $g_0$. The numerical results (solid lines) are compared against the analytical prediction from Eq.~(\ref{Eq:RMSD_Gaussian}) (black dashed line). For $g_0 \le (20\,\text{meV})^2$, the initial ballistic expansion is insensitive to the diagonal disorder, whereas very large disorder (purple line) leads to an early departure from the quadratic growth.  } \label{Fig:Gaussian_Exc_g0_Inv}
\end{figure}

Figure~\ref{Fig:Gaussian_Exc_g0_Inv} illustrates the numerical evolution of the RMSD for various on-site disorder strengths $g_0$ ranging from $0$ to $(100\,\text{meV})^2$, with fixed parameters $J = -10\, \text{meV}$, $g_1 = (10\, \text{meV})^2$, and $w_0 = 40$. The simulations show excellent agreement with the analytical prediction (black dashed line) for the first $\sim 20$ fs, provided $g_0 \le (20\,\text{meV})^2$. This agreement confirms that diagonal disorder does not influence the initial ballistic spreading in the weak-to-moderate disorder regime. Beyond this regime, the curves diverge as the magnitude of $g_0$ dictates the transition toward diffusive transport. Notably, for the extremely large disorder case $g_0 = 100 g_1$ (purple curve), the RMSD deviates earlier ($t\sim 12\,\text{fs}$), exhibiting a diffusive behavior near $t>15\,\text{fs}$ that falls outside the validity of the approximation in Eq.~(\ref{Eq:RMSD_Gaussian}).

\subsubsection{Dependence of short-time RMSD dynamics on initial wavepacket width $w_0$}

We now turn our attention to the crossover of the short-time RMSD scaling from the wide Gaussian wavepacket limit to the localized excitation limit.
Note that the validity of Eq.~(\ref{Eq:RMSD_Gaussian}) depends not only on the magnitudes of $J$, $g_0$, and $g_1$, but also on the continuum approximation, which requires $w_0$ to be sufficiently large. In the opposite limit, as $w_0\to 0$, the initial density matrix in Eq.~(\ref{Eq:Gaussian_initial_density_matrix}) reduces to the localized initial condition $\rho_{l,r}(0) = \delta_{l,0}\delta_{r,0}$. Consequently, the short-time behavior of the RMSD transitions from a quadratic dependence ($\propto t^2$) for a Gaussian wavepacket to a linear dependence ($\propto t$) for a localized excitation. To clearly illustrate this transition, we define the quantity $\Delta \text{RMSD} = \text{RMSD}(t) - \text{RMSD}(0)$, representing the growth in the mean width of the wavepacket. The analytical expressions for $\Delta \text{RMSD}$ in these two limiting cases are:
\begin{align}
\label{Eq:Delta_x_Gaussian}
\Delta \text{RMSD}  &= a \frac{ \sqrt{2} g_1 }{ w_0 \hbar^2 } t^2, \quad \text{(Wide Gaussian wavepacket)} \\
\Delta \text{RMSD}  &= a \frac{\sqrt{2(J^2+g_1)}}{\hbar} t. \quad \text{(Localized excitation)}
\label{Eq:Delta_x_Local}
\end{align}

\begin{figure}
   \includegraphics[width=0.98\linewidth]{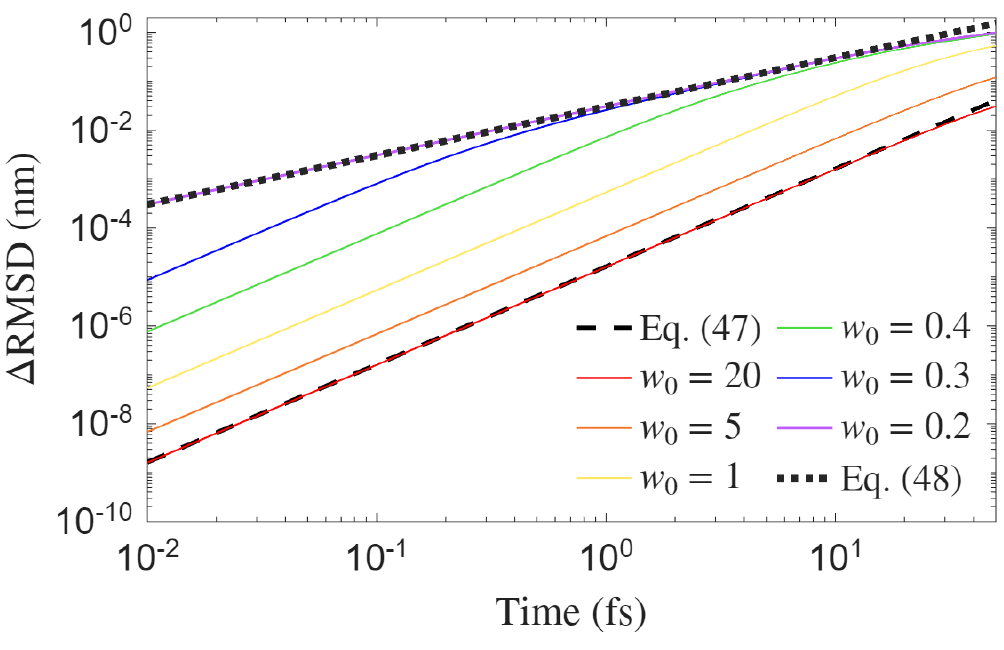}
    \caption{ Short-time evolution of $\Delta \text{RMSD}$ for various initial widths $w_0$. This log-log plot illustrates the transition from quadratic scaling ($\propto t^2$) for a broad Gaussian wavepacket ($w_0=20$) to linear scaling ($\propto t$) for a localized excitation ($w_0=0.2$). Numerical results (solid lines) show excellent agreement with the analytical limits of Eq.~(\ref{Eq:Delta_x_Gaussian}) (black dashed line) and Eq.~(\ref{Eq:Delta_x_Local}) (black dotted line). } \label{Fig:Gaussian_Exc_w0_Depd}
\end{figure}

In Figure~\ref{Fig:Gaussian_Exc_w0_Depd}, we show the time evolution of $\Delta \text{RMSD}$ on a log-log scale for initial widths ranging from $w_0 = 20$ to $w_0 = 0.2$, with the fixed parameters $J=-10~\text{meV}$, $g_1 = (10~\text{meV})^2$, $g_0=k_\parallel=0$. We observe the transition between the two scaling regimes through the change in the slope of the curves. For large initial widths ($w_0 = 20$), the numerical results perfectly overlap with the analytical prediction of Eq. (\ref{Eq:Delta_x_Gaussian}) (black dashed line), exhibiting a slope of $\sim 2$ characteristic of quadratic growth. As $w_0$ decreases, the wavepacket's behavior deviates from this continuum-limit scaling. In the highly localized limit ($w_0 = 0.2$), the growth becomes linear, as indicated by the convergence of the numerical data with Eq. (\ref{Eq:Delta_x_Local}) (black dotted line), which shows a slope of $\sim 1$. This crossover highlights how the discrete nature of the lattice begins to dominate the transport dynamics as the wavepacket's spatial extent becomes comparable to the lattice constant.

\section{Short-time exciton dynamics in 1D molecular chain with orientational disorder}
\label{Sec:MQED_Example}
With the confirmed agreement between the derived analytical expression and the numerical results, we now shift our focus to investigating the impact of disorder on exciton dynamics. Specifically, instead of directly sampling $\beta_{mn}$ with a given disorder parameter, we explicitly model the off-diagonal disorder as arising from random dipole orientations within the molecular aggregate. To do so, we first parameterize the dipole-dipole interactions (DDI) between molecular emitters using the macroscopic quantum electrodynamics (MQED) framework.

\subsection{DDI of the MQED Hamiltonian with orientational disorder}
Based on the MQED framework in the weak light-matter coupling regime, exciton dynamics in molecular aggregates can be described by the Lindbladian master equation \cite{chuang_2024_2,wang_robust_2025}:
\begin{align}
\nonumber
&\frac{\partial}{\partial t} \hat{\rho}(t) = 
-\frac{i}{\hbar} \left[ \hat{H}_{\mathrm{M}} + \hat{{H}}_\mathrm{DDI}, \hat{\rho}(t) \right] \\
&\qquad + \sum_{m,n}^{N_\mathrm{lattice}}  {\Gamma}_{mn} \left( \hat{\sigma}_{n}^{-} \hat{\rho}(t) \hat{\sigma}_{m}^{+} - \frac{1}{2} \left\{ \hat{\sigma}_{m}^{+} \hat{\sigma}_{n}^{-}, \hat{\rho}(t) \right\} \right),
\label{Eq:QME_MQED}
\end{align}
where $\hat{\sigma}^{+}_m = |m\rangle\langle \mathrm{G}|$ and $\hat{\sigma}^{-}_m = |\mathrm{G}\rangle\langle m|$ create and annihilate an exciton on the $m$-th molecule, respectively, and $|\mathrm{G}\rangle$ denotes the global ground state. The molecular Hamiltonian $\hat{H}_{\mathrm{M}} = \sum_{m}^{N_\mathrm{lattice}} E_m \hat{\sigma}^{+}_m \hat{\sigma}^{-}_m$ incorporates the on-site energies $E_m = \hbar\omega_\mathrm{M} + \Lambda^\mathrm{Sc}_m$, where $\omega_\mathrm{M}$ is the molecular transition frequency in vacuum, and $\Lambda^\mathrm{Sc}_m$ is the Casimir-Polder (CP) potential induced by the dielectric environment. The dipole-dipole interaction (DDI) is given by $\hat{{H}}_\mathrm{DDI} = \sum_{m \neq n} V_{mn} \hat{\sigma}^{+}_m \hat{\sigma}^{-}_n$. 

The CP potential $\Lambda^\mathrm{Sc}_m$, the DDI strength $V_{mn}$, and the generalized dissipation rates $\Gamma_{mn}$ are determined by the dyadic Green's function $\overline{\overline{\mathbf{G}}}(\mathbf{r}_m,\mathbf{r}_n,\omega)$ as follows:
\begin{align}
\nonumber
    &\Lambda_{m}^\mathrm{Sc} = \mathcal{P} \int_{0}^{\infty} d\omega \frac{\omega^2}{\pi \varepsilon_0 c^2}   \left( \frac{1}{ \omega + \omega_\mathrm{M} } -\frac{1}{\omega - \omega_\mathrm{M} } \right) \\
    & \qquad \quad \times {\boldsymbol{\mu}_{m} \cdot \mathrm{Im}\overline{\overline{\mathbf{G}}}_{\text{Sc}}(\mathbf{r}_m,\mathbf{r}_m,\omega) \cdot \boldsymbol{\mu}_{ m}} , 
\label{Eq:CP_Potential_MQED}
\\
\label{Eq:DDI_Strength_MQED}
    &V_{mn} =  \frac{- \omega_\mathrm{M}^2}{ \epsilon_0 c^2}  \boldsymbol{\mu}_{m} \cdot \mathrm{Re} \overline{\overline{\mathbf{G}}}(\mathbf{r}_m,\mathbf{r}_n,\omega_\mathrm{M}) \cdot \boldsymbol{\mu}_{n}, \\
    &\Gamma_{mn} = \frac{2\omega_{\mathrm{M}}^2}{\hbar\varepsilon_0 c^2} \boldsymbol{\mu}_{m} \cdot \mathrm{Im}\overline{\overline{\mathbf{G}}}(\mathbf{r}_{m},\mathbf{r}_{n},\omega_{\mathrm{M}}) \cdot \boldsymbol{\mu}_{n}. 
    \label{Eq:Interaction_Parameters_Definition}
\end{align}
Here, $\mathcal{P}$ denotes the Cauchy principal value, $c$ is the speed of light in vacuum, and $\epsilon_0$ is the vacuum permittivity. The total Green's function, decomposed into vacuum and scattering components as $\overline{\overline{\mathbf{G}}} = \overline{\overline{\mathbf{G}}}_0 + \overline{\overline{\mathbf{G}}}_{\text{Sc}}$, satisfies Maxwell's macroscopic equation\cite{scheel2008macroscopic,ding2017plasmon,hsu2017plasmon}:
\begin{align}
\left( \frac{\omega^2}{c^2}\epsilon_\mathrm{r}(\mathbf{r}_m,\omega) - \boldsymbol{\nabla} \times \boldsymbol{\nabla} \times \right)\overline{\overline{\mathbf{G}}}(\mathbf{r}_m,\mathbf{r}_n,\omega) = 
- \delta(\mathbf{r}_m-\mathbf{r}_n) \mathbf{\overline{\overline{I}}}_3,
\label{Eq:Green_Fun_full}
\end{align}
where $\epsilon_\mathrm{r}(\mathbf{r},\omega)$ is the relative permittivity and $\mathbf{\overline{\overline{I}}}_3$ is the $3\times 3$ identity matrix.

\begin{figure}
   \includegraphics[width=1.0\linewidth]{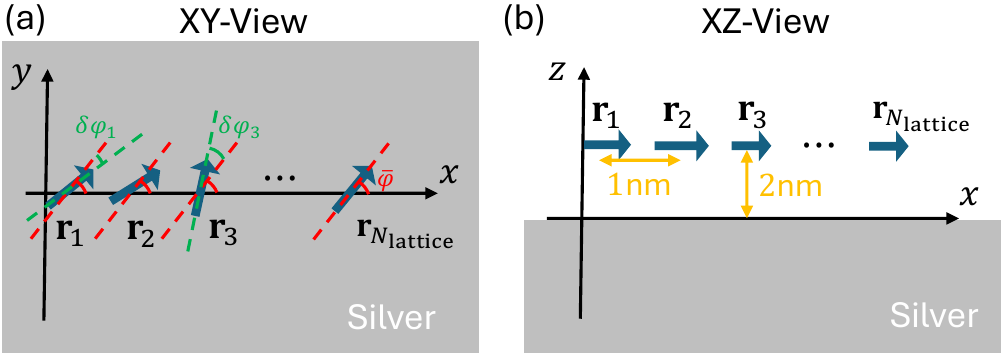}
    \caption{ Schematic representation of an array of parallel molecular dipoles above a silver plane. (a) XY-View (Top-down): The blue arrows indicate the orientations of molecular dipoles with disordered azimuthal angles. The mean azimuthal angle for the ensemble is denoted by $\bar{\varphi}$, while $\delta\varphi_m$ represents a Gaussian-distributed random fluctuation for the $m$-th dipole.  
(b) XZ-View (Side view): The dipoles are located at positions $\mathbf{r}_m$ (for $m=1, \cdots, N_{\text{lattice}}$) at a constant height of $2\,\text{nm}$ above the silver surface. All dipoles are oriented in-plane with a polar angle of $\theta = 90^\circ$ relative to the z-axis. The intermolecular spacing between adjacent dipoles is $a=1\,\text{nm}$.
} 
  \label{Fig:Schematic} 
\end{figure}

To investigate the effect of orientational disorder on exciton transport, we consider a linear chain of $N_\mathrm{lattice} = 200$ identical molecules positioned at $\mathbf{r}_m = (ma, 0, 2\,\text{nm})$ above a silver surface, with a lattice constant of $a = 1\,\text{nm}$. The dielectric environment is modeled as a planar interface:
\begin{align}
    \epsilon_\mathrm{r}(\mathbf{r},\omega) = 
    \begin{cases}
        1,     & \text{ if } z>0, \\
         \epsilon_{\mathrm{Ag}}(\omega),    & \text{ if } z<0,
    \end{cases} 
    \label{dielectrics}
\end{align}
where the silver permittivity $\epsilon_{\mathrm{Ag}}(\omega)$ is obtained from experimental data\cite{Johnson1972}.  Translational symmetry ensures a uniform on-site energy $E_m$ across the lattice. Since a constant energy shift does not affect the dynamics of the spatial moments, we set $E_m = 0$, which is equivalent to transforming the system into the interaction picture. Notably, for a dipole oriented parallel to the silver surface, azimuthal angle disorder does not induce on-site disorder $g_0$. In contrast, for a vertical or tilted dipole, polar angle disorder leads to significant on-site disorder (see Appendix~\ref{Appendix:CP_Potential} for the details). Therefore, for simplicity, this study focuses exclusively on parallel dipoles (as shown in Fig.~\ref{Fig:Schematic}) to avoid the complications introduced by $g_0$.

We model the orientational disorder by expressing the transition dipole moment of the $m$-th molecule as $\boldsymbol{\mu}_m = |\boldsymbol{\mu}_m| (\cos({\bar{\varphi}} + {\delta{\varphi_m}}), \sin({\bar{\varphi}} + {\delta{\varphi_m}}), 0)$, where $ {\bar{\varphi}}$ is the mean azimuthal angle and ${\delta{\varphi_m}}$ represents a Gaussian-distributed random fluctuation with $\langle {\delta{\varphi_m}} \rangle_\mathrm{E} = 0$. For small angular disorder ($\sigma_{\varphi} \equiv \sqrt{\langle {\delta{\varphi_m}}^2 \rangle_\mathrm{E}} \ll 1$), the Taylor expansion of Eq.~(\ref{Eq:DDI_Strength_MQED}) yields:
\begin{align}
    V_{mn} \approx V_{mn}^{[00]} + {\delta{\varphi_m}} V_{mn}^{[10]} + {\delta{\varphi_n}} V_{mn}^{[01]} ,
\label{Eq:MQED_V_approx}
\end{align}
where the coefficients are:
\begin{align}
    \nonumber
    &V_{mn}^{[00]} = \frac{- \omega_\mathrm{M}^2 |\boldsymbol{\mu}_{m}| |\boldsymbol{\mu}_{n}| }{ \epsilon_0 c^2}  \mathbf{e}_m^{[0]} \cdot \mathrm{Re} \overline{\overline{\mathbf{G}}}(\mathbf{r}_m,\mathbf{r}_n,\omega_\mathrm{M}) \cdot \mathbf{e}_n^{[0]} , \\
    \nonumber
    &V_{mn}^{[10]} =  \frac{- \omega_\mathrm{M}^2 |\boldsymbol{\mu}_{m}| |\boldsymbol{\mu}_{n}| }{ \epsilon_0 c^2}  \mathbf{e}_m^{[1]} \cdot \mathrm{Re} \overline{\overline{\mathbf{G}}}(\mathbf{r}_m,\mathbf{r}_n,\omega_\mathrm{M}) \cdot \mathbf{e}_n^{[0]} , \\
    &V_{mn}^{[01]} =  \frac{- \omega_\mathrm{M}^2 |\boldsymbol{\mu}_{m}| |\boldsymbol{\mu}_{n}| }{ \epsilon_0 c^2}  \mathbf{e}_m^{[0]} \cdot \mathrm{Re} \overline{\overline{\mathbf{G}}}(\mathbf{r}_m,\mathbf{r}_n,\omega_\mathrm{M}) \cdot \mathbf{e}_n^{[1]}.
\label{Eq:V_Approx_MQED}
\end{align} 
Here, the unit vectors $\mathbf{e}^{[0]}_{m} = (\cos\varphi_m, \sin\varphi_m, 0)$ and $\mathbf{e}^{[1]}_{m} = (-\sin\varphi_m, \cos\varphi_m, 0)$ represent the mean orientation and its orthogonal perturbation, respectively. The Onsager reciprocity theorem\cite{buhmann2013dispersion} ensures $\mathbf{e}_{\alpha} \cdot \overline{\overline{\mathbf{G}}}(\mathbf{r}_\alpha,\mathbf{r}_\beta,\omega_\mathrm{M}) \cdot \mathbf{e}_{\beta} = \mathbf{e}_{\beta} \cdot \overline{\overline{\mathbf{G}}}(\mathbf{r}_\beta,\mathbf{r}_\alpha,\omega_\mathrm{M}) \cdot \mathbf{e}_{\alpha}$, which guarantees $V_{mn} = V_{nm}$.

Having established the expansion for the orientationally disordered DDI, we now map these MQED-derived quantities onto our generic disorder model in Sec.~\ref{Sec_Hamil_Dyna}. In Eq.~(\ref{Eq:MQED_V_approx}), $V_{mn}^{[00]}$ represents the deterministic coupling strength corresponding to $\alpha_{mn}$ introduced in Eq.~(\ref{Eq:Hamiltonian}). The off-diagonal disorder is therefore $\beta_{mn} = V_{mn} - V_{mn}^{[00]}$. Since the fluctuations $ {\delta{\varphi_m}}$ and $ {\delta{\varphi_n}}$ are assumed to be statistically independent (i.e. $\langle {\delta{\varphi_m}} {\delta{\varphi_n}} \rangle_\mathrm{E} = 0$), the covariance follows directly from Eq.~(\ref{Eq:MQED_V_approx}):
\begin{align}
    \langle \beta_{mn}^2  \rangle_\mathrm{E} = \langle \delta\varphi_m^2  \rangle_\mathrm{E} \left(V_{mn}^{[10]}\right)^2 + \langle \delta\varphi_n^2  \rangle_\mathrm{E} \left(V_{mn}^{[01]}\right)^2.
\label{Eq:MQED_beta_approx}
\end{align}
In the end, we assume uniform angular disorder strength $\sigma_{\varphi} = \sqrt{ \langle {\delta{\varphi_m}}^2 \rangle_\mathrm{E} }$ for all $m$, and simplify Eq.~(\ref{Eq:MQED_beta_approx}) to the form of the static disorder strength $g(m-n)$ defined in Eq.~(\ref{Eq:Covariance_beta}):
\begin{align}
    g(m-n)= \left[ \left(V_{mn}^{[10]}\right)^2 + \left(V_{mn}^{[01]}\right)^2 \right] \sigma_{\varphi}^2.
\label{Eq:g_mn_sigma_Delta}
\end{align}

\subsection{Short-time RMSD: Robust prediction based on the effective $g_1$ }
\begin{figure}
   \includegraphics[width=0.98\linewidth]{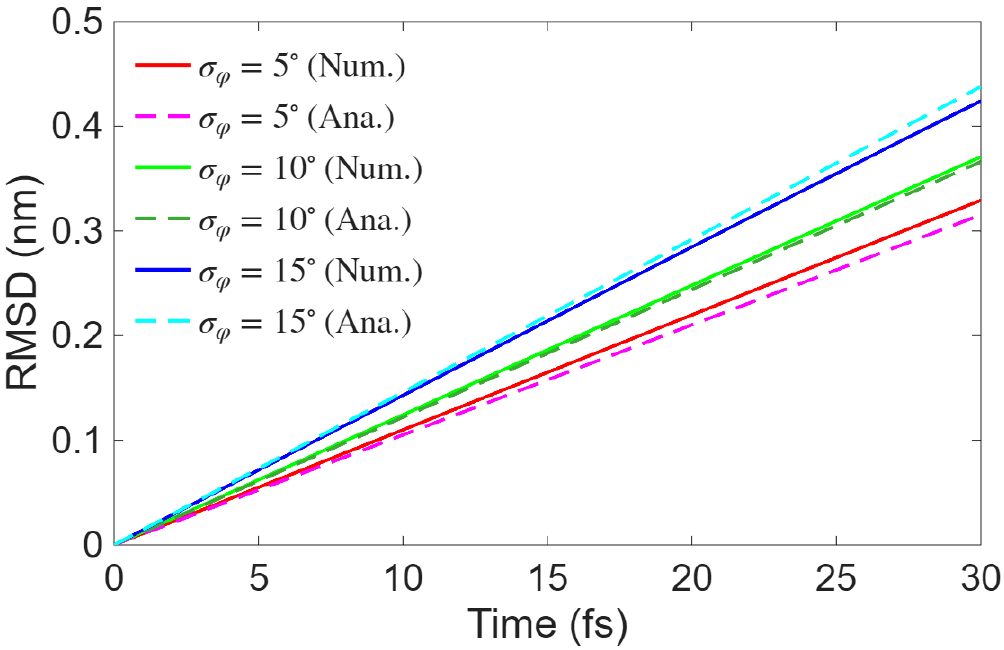}
    \caption{ Short-time RMSD for a localized initial excitation in a 1D molecular chain ($N_\text{lattice}=200$, $a=1\text{ nm}$) positioned $2\text{ nm}$ above a silver surface. The solid lines represent full numerical Lindbladian dynamics evaluated via MQED, while dashed lines show the analytical predictions based on the nearest-neighbor approximation [Eq.~(\ref{Eq:RMSD_linear})]. Dynamics are plotted for orientational disorder strengths $\sigma_{\varphi}$ of $5^\circ$ (red), $10^\circ$ (green), and $15^\circ$ (blue).  } 
    \label{Fig:MQED_Single_Exc_Compare}
\end{figure}

Figure~\ref{Fig:MQED_Single_Exc_Compare} illustrates the short-time RMSD dynamics for a localized initial excitation across varying orientational disorder strengths ($\sigma_\varphi$, from $5^\circ$ to $15^\circ$), with results of RMSD averaged over $4000$ independent trajectories. Other physical parameters are held constant: $\bar{\varphi} = 45^\circ$, $|\boldsymbol{\mu}_m| = 3.8$ Debye, and $\hbar \omega_\mathrm{M} = 1.864$ eV for evaluating $V_{mn}$ and $\Gamma_{mn}$ in Eqs.~(\ref{Eq:DDI_Strength_MQED}) and (\ref{Eq:Interaction_Parameters_Definition}). The dyadic Green's function for the silver surface Eq.~(\ref{Eq:Green_Fun_full}) is computed using the Fresnel formulation and Sommerfeld integrals\cite{wu2018characteristic}, and the RMSD is calculated by simulating the Lindbladian quantum master equation using the MQED-QD package\cite{liu2026mqed}.
The analytical predictions (dashed lines) are generated using Eq.~(\ref{Eq:RMSD_linear}), mapping the deterministic coupling $J = V_{m,m+1}^{[00]}$ and disorder strength $g_1$ according to Eqs.~(\ref{Eq:V_Approx_MQED}) and (\ref{Eq:g_mn_sigma_Delta}), respectively (see Appendix~\ref{Appendix:MQED_V_Distribution} for details).

The close agreement between the full MQED-based Lindbladian dynamics (solid lines) and our analytical model demonstrates that: (\textit{i}) the generalized dissipation rates $\Gamma_{mn}$ are negligible in this ultrafast, sub-30 fs regime; (\textit{ii}) for a small lattice constant $a=1\,\text{nm}$, the exciton dynamics are dominated by nearest-neighbor coupling and its associated static disorder, both of which are accurately captured by the analytical model; and (\textit{iii}) although the full DDI Hamiltonian (which includes long-range interactions) yields a slightly larger RMSD than the nearest-neighbor approximation (visible at $\sigma_{\varphi} = 5^\circ$ and $10^\circ$), the analytical solution performs robustly in predicting transport behavior for moderate orientational disorder ($\sigma_{\varphi} \approx 10^\circ$).

\section{Conclusions}
\label{Sec:Conclusion}

In this study, we focus on contrasting the effects of on-site energetic disorder and coupling fluctuations on the short-time dynamics of exciton transport. Leveraging reciprocal-space analysis of the LvN equation and the short-time approximation, we derive the leading-order analytical expression for the spatial moments $\langle x^n \rangle$, specifically for the localized excitation initial condition, Eq.~\eqref{Eq:x_square_Simple}, and the moving Gaussian wavepacket, Eqs.~\eqref{Eq:x_t_Gaussian_placeholder} and \eqref{Eq:x2_t_Gaussian_placeholder}. These approximate expressions delineate the distinct roles of deterministic coupling and random fluctuations in shaping the initial ballistic spread.

Our analysis of the generic disorder model reveals several interesting phenomena. 
First, while it is known that on-site energetic disorder governs the asymptotic exciton localization, it does not play a major role at short times. Instead, off-diagonal coupling fluctuations exert a dominant influence on the short-time velocity and spatial expansion of excitons. This insensitivity of the initial transport to diagonal disorder $g_0$ is demonstrated for both the localized excitation and Gaussian wavepacket initial conditions.
Second, we find that the coherent coupling $J$ and off-diagonal fluctuations $g_1$ have an equivalent contribution to the short-time ballistic transport. The effective coefficient of the $t^2$ term in $\langle x^2(t)\rangle$ is $\propto J^2+g_1$ for localized excitation and $\propto 2J^2\sin{k_\parallel}+g_1$ for a moving Gaussian wavepacket. This contribution of the disorder effect resembles the effective Rabi splitting for explaining disorder-induced spectral splitting.\cite{li_disorder-induced_2025} More importantly, our numerical results indicate that increased on-diagonal disorder $g_0$ reduces the duration of this ballistic regime, facilitating the crossover to diffusive transport (see Figs.~\ref{Fig:Local_Exc_g0_Inv} and \ref{Fig:Gaussian_Exc_g0_Inv}). 

Beyond the generic disorder model, we further apply our analysis to a more realistic system of molecular aggregates with orientational disorder. We utilize the MQED framework and estimate the key parameters, $J$ and $g_1$, based on the simulation. With the two parameters, we can capture the quantitatively accurate short-time dynamics and disorder-induced transport enhancement, which further demonstrate the robustness of our analysis. This agreement provides a conceptual bridge between generic disorder model analysis and parameter-free MQED simulation of experimentally realizable systems, such as molecular aggregates on metal surfaces.

Although our analytical treatment has provided significant insight into the short-time dynamics of exciton transport, several important directions deserve for future exploration. Having established a clear understanding of the ballistic regime, the natural next step is to develop theoretical frameworks that transcend the short-time approximation and capture the full temporal evolution of exciton wavepackets. Bridging the gap between early-time ballistic motion and the long-time regime (where phenomena such as localization or diffusive transport typically emerge) would offer a more complete and unified picture of energy propagation in disordered systems. In addition, extending the present formalism to higher-dimensional lattices and incorporating the effects of stochastic disorder (where $\beta_{mn}$ explicitly depends on time) arising from thermal and phonon fluctuations will be crucial for capturing the complexity of realistic material environments.\cite{engelhardt_unusual_2022,catuto_interplay_2025} Such advances will not only deepen our fundamental understanding of exciton dynamics but also pave the way for predictive modeling of complex molecular architectures, ultimately enabling the design of next-generation optoelectronic devices with optimized energy flow across all relevant timescales.

\begin{acknowledgments}
Wang, Liu, and Chen thank the support provided by the University of Notre Dame and the Asia Research Collaboration grant of Notre Dame Global.
\end{acknowledgments}


\section*{data availability}
The data that support the findings of this study are available from the corresponding author upon reasonable request.

\begin{appendix}

\section{Laplace transform to Eq.~(\ref{Eq:Integro-Differ})}
\label{Appendix:Laplace_Tran_Integro-Differ}
By performing the Laplace transformation to Eq.~(\ref{Eq:Integro-Differ}), and the applying identities $\mathcal{L}\left[ \partial \tilde{f}(t)/\partial t \right] = p \tilde{f}(p) - \tilde{f}(t=0) $, $\mathcal{L}\left[ \int_0^t \tilde{f}(t-\tau) \tilde{g}(\tau) d\tau \right] =  \tilde{f}(p) \tilde{g}(p) $, and $\mathcal{L}\left[ e^{iAt}  \right] = (p-iA)^{-1} $, we obtain:
\begin{align}
\nonumber
    &p \langle \tilde{\rho}(k_1,k_2;p) \rangle_\mathrm{E} = 
    \langle \tilde{\rho}(k_1,k_2;t=0) \rangle_\mathrm{E}  + \frac{i4J}{\hbar} \sin \frac{k_1+k_2}{2}\sin \frac{k_1-k_2}{2} \\
\nonumber
    & \quad \times \langle \tilde{\rho}(k_1,k_2;p) \rangle_\mathrm{E} - \frac{1}{2\pi\hbar^2} \int_{-\pi}^{\pi} dq \Bigg\{ \Big[p-\frac{i4J}{\hbar}\sin\frac{q+k_1+k_2}{2} \\
\nonumber
    & \quad \times \sin\frac{q+k_1-k_2}{2}  \Big]^{-1} \Big[ \mathcal{G}(k_1, k_1, q) \langle \tilde{\rho}(k_1,k_2;p) \rangle_\mathrm{E} - \mathcal{G}(k_1, k_2, q) \\
\nonumber
    & \quad \times \langle \tilde{\rho}\big(q+k_1,q+k_2;p\big) \rangle_\mathrm{E} 
    \Big] - \Big[p-\frac{i4J}{\hbar}\sin\frac{q+k_1+k_2}{2} \\
\nonumber
    & \quad \times \sin\frac{-q+k_1-k_2}{2}  \Big]^{-1} \Big[ \mathcal{G}(k_1, k_2, q) \langle \tilde{\rho}\left(q+k_1,q+k_2;p\right) \rangle_\mathrm{E} \\
    &\quad - \mathcal{G}(k_2, k_2, q) \langle \tilde{\rho}(k_1,k_2;p) \rangle_\mathrm{E} 
\Big] \Bigg\} .
\label{Eq:Append_Laplace_Integro-Differ_Ori}
\end{align}

By introducing the variable substitution $F(u,s;p) \equiv \langle \tilde{\rho}(k_1, k_2; p) \rangle_\mathrm{E} = F\left(k_1-k_2,\frac{k_1+k_2}{2};p\right)$, where $u = k_1 - k_2$ and $s = (k_1 + k_2)/2$, Eq.~(\ref{Eq:Append_Laplace_Integro-Differ_Ori}) is recast as:
\begin{align}
\nonumber
    &p F(u,s;p) = 
    \langle \tilde{\rho}(k_1,k_2;t=0) \rangle_\mathrm{E}  + \frac{i4J}{\hbar} \sin (s) \sin \left(\frac{u}{2} \right) F(u,s;p) \\
\nonumber
    & \quad  - \frac{1}{2\pi\hbar^2} \int_{-\pi}^{\pi} dq \Bigg\{ \Big[p-\frac{i4J}{\hbar}\sin\frac{q+2s}{2}  \sin\frac{q+u}{2}  \Big]^{-1} \\
\nonumber
    & \quad \times  \Big[ {G}(0, s+u/2, q) F(u,s;p)  - {G}(u, s, q) F(u,s+q;p)\Big]  \\
\nonumber
    & \quad    
     - \Big[p-\frac{i4J}{\hbar}\sin\frac{q+2s}{2} \sin\frac{-q+u}{2}  \Big]^{-1} \\
    & \quad \times  \Big[ {G}(u, s, q) F(u,s+q;p) - {G}(0, s-u/2, q) F(u,s;p) 
\Big] \Bigg\} ,
\label{Eq:Append_Laplace_Integro-Differ_us}
\end{align}
where the reciprocal-space disorder kernel ${G}(u, s, q) = \mathcal{G}(k_1-k_2, \frac{k_1+k_2}{2}, q)$ is rewritten as:
\begin{align}
{G}(u, s, q) = g_0 + 2g_1 \left[ \cos(q + 2s) + \cos(u) \right].
\label{Eq:Reciprocal-space disorder_kernel_us}
\end{align}

\section{Derivation of the site representation of $\Omega_1(u;p)$}
\label{Appendix:Site_representation_Omega_1}
We begin with the definition of $\Omega_1$ as the integral of the function $F(u,s;p)$. This function corresponds to the ensemble-averaged density matrix in reciprocal space, $\langle \tilde{\rho}(k_1, k_2; p) \rangle_\mathrm{E}$:
\begin{align}
\nonumber
    \Omega_1(u,p) &= \frac{1}{2\pi} \int_{-\pi}^{\pi} ds\, F(u,s,p)
    = \frac{1}{2\pi} \int_{-\pi}^{\pi} ds\, \langle \tilde{\rho}(k_1, k_2; p) \rangle_\mathrm{E}.
\end{align}

To express $\Omega_1$ in the site representation, we substitute the Fourier transform of $\langle \tilde{\rho}(k_1, k_2; p) \rangle_\mathrm{E}$ in terms of the lattice-site density matrix elements. Using the identities $k_1 = s + u/2$ and $k_2 = s - u/2$, we obtain
\begin{align}
\nonumber
    \Omega_1(u,p) &=  \frac{1}{2\pi} \int_{-\pi}^{\pi} ds \sum_{l,r=-\infty}^{+\infty} e^{-i (s+u/2) l + i (s-u/2) r}  \langle {\rho}_{l,r}(p) \rangle_\mathrm{E} \\
    & = \frac{1}{2\pi} \int_{-\pi}^{\pi} ds \sum_{l,r=-\infty}^{+\infty}  e^{-i s( l-r) - i \frac{u}{2}( l+r) } \langle {\rho}_{l,r}(p)\rangle_\mathrm{E}.
\label{Eq:Append_Site_Rep_Omega1}
\end{align}

Equation~(\ref{Eq:Append_Site_Rep_Omega1}) can now be evaluated by carrying out the integral over $s$. Specifically, the integration of the phase factor $e^{-is(l-r)}$ over the interval $[-\pi,\pi]$ yields $2\pi$ times the Kronecker delta $\delta_{l,r}$. Applying this orthogonality relation eliminates the integral and gives
\begin{align}
     \Omega_1(u;p) = \sum_{l,r=-\infty}^{+\infty} \delta_{l,r}\, e^{- i \frac{u}{2}( l+r) } \langle {\rho}_{l,r}(p)  \rangle_\mathrm{E}.
\end{align}

Finally, the Kronecker delta collapses the double summation, since only terms with $l=r$ contribute. Therefore, $\Omega_1$ reduces to the Fourier transform of the diagonal density-matrix elements, namely the site populations, in the Laplace domain:
\begin{align}
    \Omega_1(u;p) =   \sum_{l=-\infty}^{+\infty}  e^{- i u l} \langle {\rho}_{l,l}(p) \rangle_\mathrm{E}.
\end{align}

\section{Explicit expression of $\Omega_1$ under local excitation initial condition}
\label{Appendix:Omega1_Local_Excit}
To obtain a closed-form expression for $\Omega_1(u;p)$, we begin with the determinant $\Delta$ defined in Eq.~(\ref{Eq:Omega_1_Formal}):
\begin{align}
\Delta = \left[1 - \frac{\Gamma_0+\Gamma_1\cos u}{p} C_1 \right] \left[ 1 + \frac{\Gamma_1}{p}C_6 \right] + \frac{\Gamma_1(\Gamma_0+\Gamma_1\cos u)}{p^2} C_3^2.
\label{Eq:Appendix_Delta_appendix_rewrite}
\end{align}

To simplify Eq.~(\ref{Eq:Appendix_Delta_appendix_rewrite}), we introduce the following parameters:
\begin{align}
R \equiv \sqrt{A^2-B^2}, \qquad
P \equiv \frac{\Gamma_0+\Gamma_1\cos u}{p}, \qquad
Q \equiv \frac{\Gamma_1}{p}.
\label{Eq:Appendix_RPQ}
\end{align}

Using the identity $B^2=(A-R)(A+R)$, the coefficients from Eqs.~(\ref{Eq:Coefficients}) and (\ref{Eq:Appendix_Delta_appendix_rewrite}) can be expressed in terms of the parameters in Eq.~(\ref{Eq:Appendix_RPQ}) as:
\begin{align}
\nonumber
C_1 &= \frac{1}{R}\\
\nonumber
C_3 &= \frac{1}{B}\left(1-\frac{A}{R}\right) = -\frac{B}{R(A+R)},
\\
C_6 &= -\frac{A}{B^2}\left(1-\frac{A}{R}\right) = \frac{A}{R(A+R)}.
\label{Eq:Appendix_C3C6_compact}
\end{align}

Substituting these into Eq.~(\ref{Eq:Appendix_Delta_appendix_rewrite}), we obtain the compact form:
\begin{align}
\Delta
&= \left(1-\frac{P}{R}\right)
\left(1+\frac{Q A}{R(A+R)}\right)
+ PQ\,\frac{B^2}{R^2(A+R)^2}
\nonumber\\
&= 1-\frac{P}{R}
+\frac{Q(A-P)}{R(A+R)}.
\label{Eq:Appendix_Delta_compact_appendix}
\end{align}

Similarly, the expression for $\Omega_1$ in Eq.~(\ref{Eq:Omega_1_Formal}) can be rewritten using Eq.~(\ref{Eq:Appendix_RPQ}):
\begin{align}
\Omega_1
=
\frac{
\begin{vmatrix}
S_1 & Q C_3 \\
S_3 & 1+Q C_6
\end{vmatrix}
}{\Delta}
=
\frac{S_1(1+Q C_6)-Q C_3 S_3}{\Delta}.
\label{Eq:Appendix_Omega1_Cramer_general}
\end{align}

For the initial condition of local excitation, where $S_1=C_1$ and $S_3=C_3$, Eq.~(\ref{Eq:Appendix_Omega1_Cramer_general}) becomes:
\begin{align}
\Omega_1
=
\frac{C_1(1+Q C_6)-Q C_3^2}{\Delta}
=
\frac{C_1+Q(C_1C_6-C_3^2)}{\Delta}.
\label{Eq:Appendix_Omega1_before_simplification}
\end{align}
Since
\begin{align}
C_1C_6-C_3^2
=
\frac{1}{R}\frac{A}{R(A+R)}
-
\frac{B^2}{R^2(A+R)^2}
=
\frac{1}{R(A+R)},
\label{Eq:Appendix_C1C6_minus_C3sq}
\end{align}
substituting Eq.~(\ref{Eq:Appendix_C1C6_minus_C3sq}) into Eq.~(\ref{Eq:Appendix_Omega1_before_simplification}) yields:
\begin{align}
\Omega_1
=
\frac{
\dfrac{1}{R}
+
\dfrac{Q}{R(A+R)}
}{\Delta}
=
\frac{A+R+Q}{R(A+R)\Delta}.
\label{Eq:Appendix_Omega1_preDelta}
\end{align}

By multiplying Eq.~(\ref{Eq:Appendix_Delta_compact_appendix}) by $R(A+R)$, we find the expression of the denominator of $\Omega_1$ in Eq.~(\ref{Eq:Appendix_Omega1_preDelta}):
\begin{align}
R(A+R)\Delta
=
(R-P)(A+R)+Q(A-P).
\label{Eq:Appendix_RARDelta}
\end{align}
Inserting Eq.~(\ref{Eq:Appendix_RARDelta}) into Eq.~(\ref{Eq:Appendix_Omega1_preDelta}) gives:
\begin{align}
\Omega_1 =\frac{A+R+Q}{(R-P)(A+R)+Q(A-P)}.
\label{Eq:Appendix_Omega1_compact_final}
\end{align}

The above result can be cast into a more compact form by introducing the auxiliary variables:
\begin{align}
X \equiv A+R+Q,
\qquad
Y \equiv A-P.
\label{Eq:Appendix_XY_definition}
\end{align}

With these definitions, the denominator in Eq.~(\ref{Eq:Appendix_Omega1_compact_final}) can be rewritten as
\begin{align}
XY-B^2
&=
(A+R+Q)(A-P)-B^2
\nonumber\\
&=
(R-P)(A+R)+Q(A-P),
\label{Eq:XY_minus_B2_identity}
\end{align}
where we have again used the identity $B^2=(A+R)(A-R)$. Consequently, $\Omega_1(u;p)$ can be expressed in the compact form:
\begin{align}
\Omega_1(u;p)=\frac{X}{XY-B^2},
\label{Eq:Appendix_Omega1_XY_form}
\end{align}
with
\begin{align}
X=A+\sqrt{A^2-B^2}+\frac{\Gamma_1}{p},
\qquad
Y=A-\frac{\Gamma_0+\Gamma_1\cos u}{p}.
\label{Eq:Appendix_XY_explicit}
\end{align}

\section{Second derivative of $\Omega_1$ for a localized initial state}
\label{Appendix_Second_Derive_Omega1_Local}
The second derivative of $\Omega_1(u;p)$ at $u=0$ is evaluated by Taylor expanding the auxiliary functions defined in Eq.~(\ref{Eq:XY_explicit}). Expanding the $u$-dependent terms $\cos u \approx 1 - u^2/2$ and $\sin^2(u/2) \approx u^2/4$ to order $O(u^2)$ yields the following expressions for $X$, $Y$, and $B^2$:
\begin{align}
X \approx X_0 + \frac{2J^2}{A\hbar^2}u^2 , \quad
Y\approx p + \frac{\Gamma_1}{2p}u^2 , \quad
B^2 \approx - \frac{4J^2}{\hbar^2} u^2 ,
\label{Eq:Appendix_XYB_u_expansion}
\end{align}
where $X_0 \equiv 2A + \Gamma_1/p$ and $A = p + (\Gamma_0+\Gamma_1)/p$. Substitution of these expansions into Eq.~(\ref{Eq:Omega1_XY_form}) results in
\begin{align}
    \Omega_1(u;p) \approx \frac{X_0 + \frac{2J^2}{A\hbar^2}u^2}{ X_0p + \left( X_0\frac{\Gamma_1}{2p} + p\frac{2J^2}{A\hbar^2} + \frac{4J^2}{\hbar^2} \right) u^2 } .
\label{Eq:Appendix_Omega_1_local_excitation_2nd_Deriv}
\end{align}

Considering a general rational function of the form $\Omega_1(u;p) = (N_0+N_1 u + N_2 u^2)/(\Delta_0 + \Delta_1 u + \Delta_2 u^2 )$, the first and second derivatives at $u=0$ are obtained via the quotient rule as:
\begin{align}
\label{Eq:Appendix_1st_2nd_Deriv_Omega_1_1st}
\left. \frac{\partial \Omega_1(u,p)}{\partial u } \right|_{u=0} &=\frac{N_1\Delta_0 - N_0\Delta_1 }{\Delta^2_0},\\ 
\left. \frac{\partial^2 \Omega_1(u,p)}{\partial u^2 } \right|_{u=0}&= \frac{ 2 \left(N_2 \Delta_0^2 - N_1\Delta_0\Delta_1 + N_0 \Delta_1^2 - N_0\Delta_0 \Delta_2 \right) }{\Delta_0^3} .
\label{Eq:Appendix_1st_2nd_Deriv_Omega_1}
\end{align}

Comparison of Eq.~(\ref{Eq:Appendix_Omega_1_local_excitation_2nd_Deriv}) with the coefficients in Eq.~(\ref{Eq:Appendix_1st_2nd_Deriv_Omega_1}) identifies the second derivative as
\begin{align}
\nonumber
\left. \frac{\partial^2 \Omega_1(u,p)}{\partial u^2 } \right|_{u=0}& =  -\frac{\Gamma_1}{p^3} - \frac{8J^2}{p^2 X_0 \hbar^2} \\
& =  -\frac{\Gamma_1}{p^3} - \frac{8J^2}{p \left( 2p^2 + 2\Gamma_0 + 3\Gamma_1 \right) \hbar^2} .
\end{align}

\section{Gaussian wavepacket initial condition in the reciprocal space}
\label{Appendix:Gaussian_initial_condition}

For the moving Gaussian wavepacket discussed in Sec.~\ref{Sec:Short-time_behavior_Observables}, the normalization condition $\sum_l |\rho_{l,l}(0)|^2=1$ implies
\begin{align}
\mathcal{N}^{-2}=\sum_{l=-\infty}^{\infty}
\exp\!\left(-\frac{l^2}{w_0^2}\right),
\label{Eq:Gaussian_norm_exact_appendix}
\end{align}
which, in the continuum approximation ($\sum_{l}\rightarrow \int dl$), reduces to
\begin{align}
\mathcal{N}
\approx
(\sqrt{\pi}\,w_0)^{-1/2}.
\label{Eq:Gaussian_norm_approx_appendix}
\end{align}

The initial state in reciprocal space, $\tilde{\rho}(k_1,k_2;0)$, is obtained by taking the Fourier transform of the site-representation density matrix:
\begin{align}
\nonumber
&\tilde{\rho}(k_1,k_2;0)
 = \sum_{l,r=-\infty}^{\infty}
e^{-ik_1 l}e^{ik_2 r}\rho_{l,r}(0)\\
& \qquad = \mathcal{N}^2
\sum_l
e^{-i(k_1-k_\parallel)l}
e^{-\frac{l^2}{2w_0^2}}
\sum_r
e^{i(k_2-k_\parallel)r}
e^{-\frac{r^2}{2w_0^2}}
\label{Eq:Gaussian_FT_def_appendix}.
\end{align}

Approximating each sum as a Gaussian integral, we have
\begin{align}
\int_{-\infty}^\infty dl
e^{-i(k_1-k_\parallel)l}
e^{-\frac{l^2}{2w_0^2} }
&\approx
\sqrt{2\pi}\,w_0
e^{-\frac{w_0^2}{2}(k_1-k_\parallel)^2 } .
\end{align}

Substituting this result into Eq.~(\ref{Eq:Gaussian_FT_def_appendix}) yields
\begin{align}
\tilde{\rho}(k_1,k_2;0)
=
2\pi w_0^2\mathcal{N}^2
e^{-\frac{w_0^2}{2}\left[(k_1-k_\parallel)^2+(k_2-k_\parallel)^2 \right] }.
\label{Eq:Gaussian_FT_k1k2_appendix}
\end{align}

Finally, by applying the normalization from Eq.~(\ref{Eq:Gaussian_norm_approx_appendix}) and transforming to the coordinates $s=\frac{k_1+k_2}{2}$ and $u=k_1-k_2$, we obtain
\begin{align}
\tilde{\rho}(u,s;0)
=
2 \sqrt{\pi} w_0 \exp\!\left[-w_0^2(s-k_\parallel)^2\right]
\exp\!\left[-\frac{w_0^2}{4}u^2\right].
\label{Eq:Gaussian_FT_us_appendix}
\end{align}

\section{Closed-form expression for $\Omega_1(u;p)$ for a moving Gaussian wavepacket}
\label{Appendix:Omega1_Gaussian}

According to the general solution derived in Eq.~(\ref{Eq:Appendix_Omega1_Cramer_general}) of Appendix~\ref{Appendix:Omega1_Local_Excit}, the integral moment $\Omega_1(u;p)$ depends on the initial conditions through the source coefficients $S_1$ and $S_3$ defined in Eq.~(\ref{Eq:Source}). For a Gaussian wavepacket, substituting Eq.~(\ref{Eq:Gaussian_FT_us_appendix}) into these coefficients yields:
\begin{align}
\nonumber
S_1 
&= \frac{1}{2\pi} \int_{-\pi}^{\pi} ds \frac{ 2 \sqrt{\pi} w_0 e^{-w_0^2(s-k_\parallel)^2}
e^{-\frac{w_0^2}{4}u^2} }{ p + \frac{2(g_0+2g_1)}{ \hbar^2 p}  -  \frac{i4J}{\hbar} \sin(s) \sin(\frac{u}{2}) } \\
&= \frac{1}{\sqrt{\pi} } \int_{-\pi}^{\pi} ds \frac{ w_0 e^{-w_0^2(s-k_\parallel)^2}
e^{-\frac{w_0^2}{4}u^2} }{ A + B \sin(s) }.
\label{Eq:CInit1_Gaussian_appendix}
\end{align}

Since the wavepacket is narrow in reciprocal space, the term $e^{-w_0^2(s-k_\parallel)^2}$ contributes significantly only near $s \approx k_\parallel$. In this limit where the normalized Gaussian behaves as a Dirac delta function, $\frac{w_0}{\sqrt{\pi}} e^{-w_0^2 (s-k_\parallel)^2 } \to \delta(s-k_\parallel)$, and Eq.~(\ref{Eq:CInit1_Gaussian_appendix}) reduces to:
\begin{align}
     S_1 \approx \frac{e^{-\frac{w_0^2}{4}u^2} }{ A + B \sin(k_\parallel) }.
\end{align}
Similarly, for $S_3$ we obtain:
\begin{align}
    S_3 \approx \frac{ \sin(k_\parallel) e^{-\frac{w_0^2}{4}u^2} }{ A + B \sin(k_\parallel) } = \sin(k_\parallel)  S_1 .
\end{align}

Consequently, the expression for $\Omega_1(u;p)$ under the Gaussian wavepacket initial condition becomes:
\begin{align}
\Omega_1(u;p) = S_1 \frac{(1+Q C_6)-Q C_3 \sin(k_\parallel)   }{\Delta}.
\label{Eq:Omega1_Gaussian_appendix}
\end{align}

By applying Eqs.~(\ref{Eq:Appendix_C3C6_compact}), (\ref{Eq:Appendix_RARDelta}), and (\ref{Eq:XY_minus_B2_identity}), we arrive at:
\begin{align}
\Omega_1(u;p) = e^{-\frac{w_0^2}{4}u^2}  \frac{ R(A+R) + Q(A+B\sin{k_\parallel})   }{ (A+B\sin{k_\parallel})(XY-B^2)  }.
\label{Eq:Appendix_Omega1_Gaussian_Near_Final}
\end{align}

Using the identity $R^2=A^2-B^2$ and the definition in Eq.~(\ref{Eq:Appendix_XY_definition}), the numerator of Eq.~(\ref{Eq:Appendix_Omega1_Gaussian_Near_Final}) can be simplified into a more compact form:
\begin{align}
    \Omega_1(u;p) = e^{-\frac{w_0^2}{4}u^2}  \frac{ AX + QB\sin{k_\parallel} -B^2   }{ (A+B\sin{k_\parallel})(XY-B^2)  }.
\label{Eq:Appendix_Omega1_Gaussian_Final}    
\end{align}

\section{First and second derivative of $\Omega_1$ for a moving Gaussian wavepacket}
\label{Appendix:Gaussian_small_u_expansion}

To evaluate the first two spatial moments, we begin by expanding Eq.~(\ref{Eq:Appendix_Omega1_Gaussian_Final}) around $u=0$. The Gaussian factor from the initial condition expands as:
\begin{align}
\exp\!\left(-\frac{w_0^2}{4}u^2\right)
=
1-\frac{w_0^2}{4}u^2+O(u^4),
\label{Eq:Gaussian_u_expansion_appendix}
\end{align}
and the $u$-dependence of the remaining coefficients $B$, $X$ and $Y$ is presented in Eq.~(\ref{Eq:Appendix_XYB_u_expansion}). Collecting terms up to second order in $u$, we obtain:
\begin{align}
\Omega_1(u;p)
= \frac{AX_0 - i\frac{2J\Gamma_1 \sin k_\parallel}{\hbar p} u + \left(\frac{6J^2}{\hbar^2}-\frac{AX_0 w_0^2}{4}\right) u^2 }{A X_0 p - i\frac{2J X_0 p \sin k_\parallel }{ \hbar } u + \left( \frac{ AX_0 \Gamma_1 }{2p} + \frac{ 2J^2 (p + 2A) }{\hbar^2}  \right) u^2 } .
\label{Eq:Omega1_Gaussian_Taylor_appendix}
\end{align}

Comparing the structure of Eq.~(\ref{Eq:Omega1_Gaussian_Taylor_appendix}) with the general expansion coefficients established in Eqs.~(\ref{Eq:Appendix_1st_2nd_Deriv_Omega_1_1st}) and (\ref{Eq:Appendix_1st_2nd_Deriv_Omega_1}) yields the first and second derivatives evaluated at $u=0$:
\begin{align}
\left.\frac{\partial \Omega_1(u;p)}{\partial u}\right|_{u=0}
=& \frac{i4J \sin k_\parallel }{ \hbar ( 2p^2 +2\Gamma_0 +3\Gamma_1 ) } , 
\label{Eq:dOmega1_du_Gaussian_appendix}
\\
\nonumber
\left.\frac{\partial^2 \Omega_1(u;p)}{\partial u^2}\right|_{u=0}
=&
\frac{ - w_0^2}{2p}
-\frac{\Gamma_1}{p^3} \\
& -\frac{8J^2\left(\Gamma_0+\Gamma_1+2p^2\sin^2 k_\parallel\right)}
{\hbar^2\,p\,(p^2+\Gamma_0+\Gamma_1)\,(2p^2+2\Gamma_0+3\Gamma_1)} .
\label{Eq:d2Omega1_du2_Gaussian_appendix}
\end{align}

Applying the spatial moment definition from Eq.~(\ref{Eq:Spatial_Moment_Def}) and expanding Eq.~(\ref{Eq:d2Omega1_du2_Gaussian_appendix}) up to order ${O}(1/p^3)$, we arrive at the Laplace-space spatial moments:
\begin{align}
\langle x(p)\rangle
&= \frac{-4aJ \sin k_\parallel }{ \hbar ( 2p^2 +2\Gamma_0 +3\Gamma_1 ) },
\\
\langle x^2(p)\rangle
&\approx a^2\left(
\frac{  w_0^2}{2p}
+\frac{ \Gamma_1}{p^3} + \frac{8J^2 \sin^2k_\parallel}{ \hbar^2 p^3 } \right) .
\end{align}

Performing an inverse Laplace transform on these expressions directly yields the short-time dynamical behaviors presented in the main text.

\section{CP potential for parallel and perpendicular orientations of a molecular emitter near a silver surface.}
\label{Appendix:CP_Potential}
The CP potential given in Eq.~(\ref{Eq:CP_Potential_MQED}) can be decomposed into two integral contributions, $\Lambda_{m}^\mathrm{Sc} = I_1 - I_2$, defined as:
\begin{align}
\label{Eq:Appendix_I1_Integr}
    &I_1 = \mathcal{P} \int_{0}^{\infty} d\omega \frac{\omega^2}{\pi \varepsilon_0 c^2}     \frac{{\boldsymbol{\mu}_{m} \cdot \mathrm{Im}\overline{\overline{\mathbf{G}}}_{\text{Sc}}(\mathbf{r}_m,\mathbf{r}_m,\omega) \cdot \boldsymbol{\mu}_{ m}}}{ \omega + \omega_\mathrm{M} } , \\
    &I_2 =  \mathcal{P} \int_{0}^{\infty} d\omega \frac{\omega^2}{\pi \varepsilon_0 c^2}     \frac{ {\boldsymbol{\mu}_{m} \cdot \mathrm{Im}\overline{\overline{\mathbf{G}}}_{\text{Sc}}(\mathbf{r}_m,\mathbf{r}_m,\omega) \cdot \boldsymbol{\mu}_{ m}} }{\omega - \omega_\mathrm{M} }  .
\label{Eq:Appendix_I2_Integr}
\end{align}

Following established contour integration techniques\cite{dzsotjan2011dipole,chuang_2024}, these integrals can be efficiently evaluated by shifting the integration path to the positive imaginary frequency axis ($\omega = i\kappa$). This transformation is mathematically advantageous because the scattering Green's function, $\overline{\overline{\mathbf{G}}}_{\text{Sc}}(\mathbf{r}_m,\mathbf{r}_m, i \kappa )$, decays rapidly and smoothly without oscillations along this axis. Evaluating the contour integrals of Eqs.~(\ref{Eq:Appendix_I1_Integr}) and (\ref{Eq:Appendix_I2_Integr}) yield:
\begin{align}
    &I_1 = - \int_{0}^{\infty} d\kappa \frac{\omega_\mathrm{M} \kappa^2 }{\pi \varepsilon_0 c^2}     \frac{{\boldsymbol{\mu}_{m} \cdot \mathrm{Re}\overline{\overline{\mathbf{G}}}_{\text{Sc}}(\mathbf{r}_m,\mathbf{r}_m, i\kappa) \cdot \boldsymbol{\mu}_{ m}}}{ \kappa^2 + \omega_\mathrm{M}^2 },\\
\nonumber
    &I_2 =  \frac{ \omega_\mathrm{M}^2 }{ \varepsilon_0 c^2}      {\boldsymbol{\mu}_{m} \cdot \mathrm{Re}\overline{\overline{\mathbf{G}}}_{\text{Sc}}(\mathbf{r}_m,\mathbf{r}_m, \omega_\mathrm{M} ) \cdot \boldsymbol{\mu}_{ m}}     \\   
    & \qquad + \int_{0}^{\infty} d\kappa \frac{ \omega_\mathrm{M} \kappa^2 }{\pi \varepsilon_0 c^2}     \frac{ {\boldsymbol{\mu}_{m} \cdot \mathrm{Re}\overline{\overline{\mathbf{G}}}_{\text{Sc}}(\mathbf{r}_m,\mathbf{r}_m, i\kappa) \cdot \boldsymbol{\mu}_{ m}} }{\kappa^2 + \omega_\mathrm{M}^2 }.
\end{align}

\begin{figure}
   \includegraphics[width=0.98\linewidth]{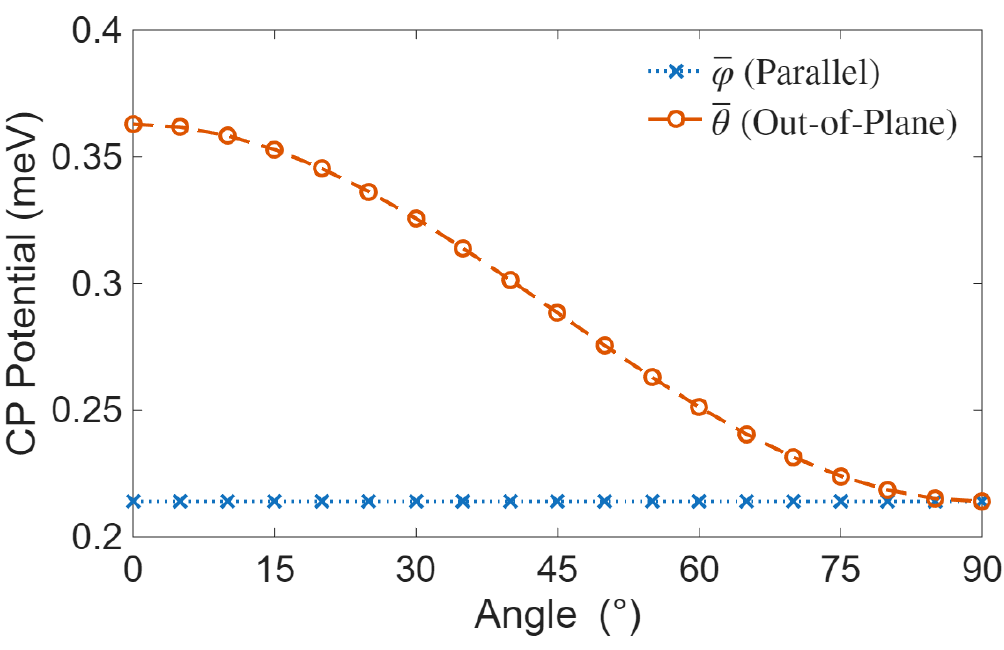}
    \caption{ CP potential $\Lambda_{m}^\mathrm{Sc}$ of a molecular emitter near a silver surface as a function of transition dipole orientations. The blue dotted line (crosses) represents a dipole oriented parallel to the surface, showing invariance with respect to the azimuthal angle $\bar{\varphi}$. The orange dashed line (circles) represents a dipole with an out-of-plane component, where the CP potential varies significantly with the polar angle $\bar{\theta}$, ranging from a maximum at $\bar{\theta} = 0^\circ$ (perpendicular to the surface) to a minimum at $\bar{\theta} = 90^\circ$ (parallel to the surface). } 
    \label{Fig:CP_Potential}
\end{figure}

In Fig.~\ref{Fig:CP_Potential}, we evaluate the CP potential $\Lambda_{m}^\mathrm{Sc}$ for two distinct transition dipole orientations. First, we consider a dipole oriented parallel to the silver surface, $\boldsymbol{\mu}_m = |\boldsymbol{\mu}_m| (\cos \bar{\varphi}, \sin \bar{\varphi}, 0)$, consistent with the model discussed in the main text. Second, we consider a dipole with an out-of-plane component, parameterized by the angle $\bar{\theta}$ as $\boldsymbol{\mu}_m = |\boldsymbol{\mu}_m| (\sin \bar{\theta}, 0, \cos\bar{\theta})$.

Evidently, for the parallel dipole, the CP potential is invariant with respect to $\bar{\varphi}$ due to the cylindrical symmetry of the dipole-plane system. Consequently, the angular disorder for parallel dipoles--as treated in the main text--does not induce diagonal disorder $g_0$. In contrast, for dipoles with an out-of-plane component, the CP potential varies significantly with $\bar{\theta}$. This suggests that angular disorder for vertical or tilted dipoles would contribute additional on-site disorder, a phenomenon that warrants further investigation in future studies.

\section{Statistical distribution of coupling strength $V_{m,m+1}$ for an orientational disordered molecular chain on silver surface}
\label{Appendix:MQED_V_Distribution}

\begin{figure}
   \includegraphics[width=0.98\linewidth]{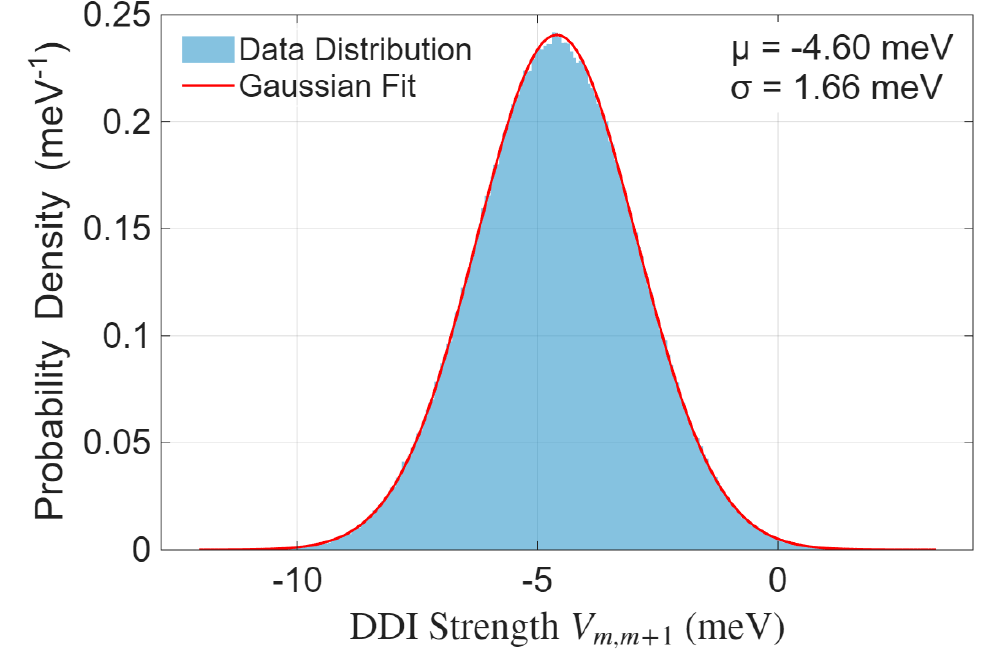}
    \caption{ Statistical distribution of the nearest-neighbor DDI strength. Probability density of $V_{m,m+1}$ for a 1D molecular chain with orientational disordered molecules situated $2\text{ nm}$ above a silver surface (with lattice constant $a = 1\text{ nm}$). The histogram (solid bars) is constructed by sampling $V_{m,m+1}$ across all sites $m \in [1, N_{\text{lattice}}]$ within the chain and over 4000 independent trajectories for $\sigma_{\varphi} = 5^\circ$. The red curve represents a Gaussian fit to the data, yielding a mean $\mu = -4.60\,\text{meV}$ and a standard deviation $\sigma = 1.66\,\text{meV}$. } \label{Fig:V_Distribution_5degree}
\end{figure}

Figure~\ref{Fig:V_Distribution_5degree} presents the statistical distribution of the nearest-neighbor DDI strength, $V_{m,m+1}$, calculated via Eq.~(\ref{Eq:DDI_Strength_MQED}) for an orientational disorder characterized by $\sigma_{\varphi} = 5^\circ = \pi/36 $. The histogram data, compiled from 4000 independent trajectories, is excellently described by a Gaussian probability density function with a mean $\mu = -4.60\,\text{meV}$ and a standard deviation $\sigma = 1.66\,\text{meV}$. Crucially, these statistical results closely match the analytical parameters derived from Eqs.~(\ref{Eq:V_Approx_MQED}) and (\ref{Eq:g_mn_sigma_Delta}), which yield a deterministic coupling $V_{m,m+1}^{[00]} \approx -4.65\,\text{meV}$ and a disorder variance $g_1 \equiv g(1) \approx (1.67\,\text{meV})^2$ for $\sigma_{\varphi} = 5^\circ$. This good agreement between $\mu\approx V_{m,m+1}^{[00]}$ and $\sigma^2 \approx g_1$ validates the mapping of the orientational fluctuations from MQED onto our generic disorder model.

\end{appendix}

\bibliography{references}

\end{document}